\documentclass[preprint,pre,amsmath,amssymb]{revtex4-1}

\usepackage{graphicx}
\usepackage{amsmath}
\usepackage{bm}
\usepackage{color}
\usepackage{dcolumn}
\usepackage{ulem}
\usepackage{hhline}
\usepackage{float}

\begin{document}

\title{Interface collisions}
\author{F. D. A. Aar\~ao Reis${}^1$ and O. Pierre-Louis${}^2$}
\affiliation{${}^1$ Instituto de F\'\i sica, Universidade Federal Fluminense,
Avenida Litor\^anea s/n, 24210-340 Niter\'oi RJ, Brazil\\
${}^2$ ILM, University  Lyon 1, 43 Bd du 11 novembre 1918, 69622 Villeurbanne, France}
\date{\today}

\begin{abstract}

We provide a theoretical framework to analyze the properties of
frontal collisions of two growing interfaces considering different short
range interactions between them.
Due to their roughness, the collision events spread in time
and form rough domain boundaries,
which defines collision interfaces in time and space.
We show that statistical properties of such interfaces depend on the kinetics of
the growing interfaces before collision, but are independent of the
details of their interaction and of their fluctuations during the collision.
Those properties exhibit dynamic scaling with exponents related to the growth
kinetics, but their distributions may be non-universal.
These results are supported by simulations of
lattice models with irreversible dynamics and local interactions.
Relations to first passage processes are discussed and a possible application to grain
boundary formation in two-dimensional materials is suggested.

\end{abstract}

\maketitle

Interface motion and collisions are ubiquitous 
in non-equilibrium systems.
For example, in graphene growth on metal substrates,
mono-crystalline domains grow and meet, ultimately forming
a polycrystalline film with grain
boundaries~\cite{Gao2010,Huang2011,Yu2011,kiraly2013}.
The formation of rough domain boundaries via interface collisions is 
encountered in many other systems undergoing domain growth, such as  
bacterial colonies~\cite{Beer2009}.
Motivated by the selection of grains in crystal growth~\cite{Saito1995} or  
of species in population dynamics~\cite{Kuhr2011},
domain boundary formation has been investigated
within competitive growth models, where two interfaces
grow in the same direction generating two types
of domains growing side by side. 
The domain boundary exhibits a self-similar behavior~\cite{Saito1995},
which can be affected by the average orientation
of the growing interfaces~\cite{Derrida1991}.
However, fewer studies have considered domain boundary formation
by frontal collisions, where colliding interfaces are parallel in average.
Based on simulations of the Eden model, Albano {\it et al}\cite{Albano1997,Albano2001} 
have exhibited numerical evidence suggesting dynamic scaling.

Furthermore, interface collisions do not always produce a domain
boundary, and instead interfaces may simply annihilate.
In such cases, the collision spreads in time
due to the roughness of the growing fronts.
This is for example observed in 
magnetic domains~\cite{KrusinElbaum2001},
reaction fronts~\cite{Atis2015}, 
turbulent liquid crystals~\cite{Takeuchi2010},
burning paper~\cite{Maunuksela1997},
forest fires~\cite{Guisoni2011},
and layer by layer crystal growth~\cite{Pimpinelli1998}. 

In this Rapid Communication, we determine both the roughness of the resulting domain
boundary, and the spreading of the collision in time during frontal collisions.
We use several different models of interface growth with irreversible rules
and short range interactions between the two interfaces.
We show that the distribution and spatial correlations 
of collision times and of the resulting domain boundary
are  independent of the details of the interactions between the two interfaces,
and only depend the roughness that builds up before collision. 
Dynamic scaling appears as a consequence of these results.  
The asymptotic distributions are dictated by the interface with the largest 
roughness when the growth exponent of the two colliding interfaces
are different, and those distributions are
non-universal when the growth exponents are equal.

\begin{figure*}
\begin{minipage}{8cm}
\begin{center}
\includegraphics[width=8cm]{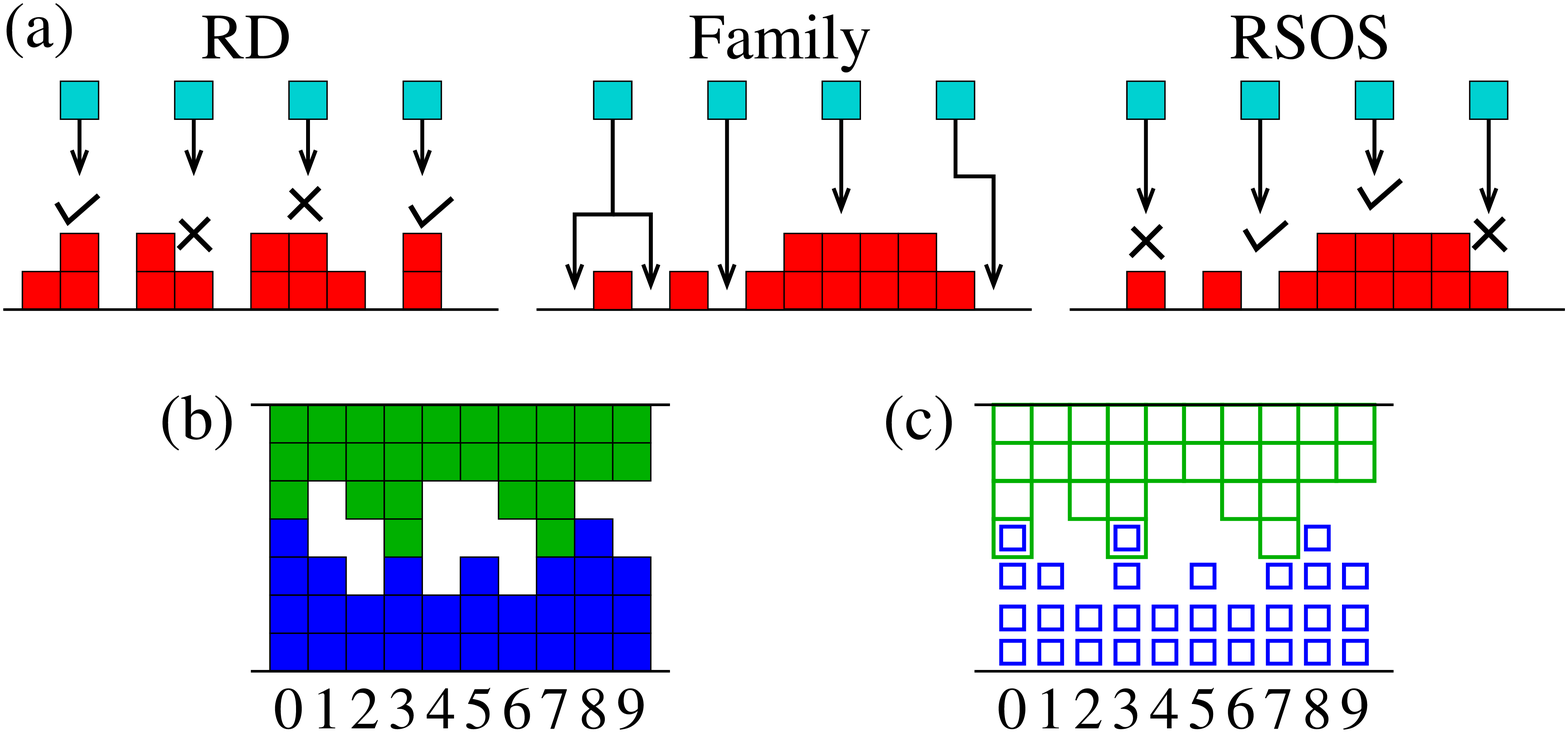}
\end{center}
\end{minipage}
\begin{minipage}{8cm}
\includegraphics[width=8cm]{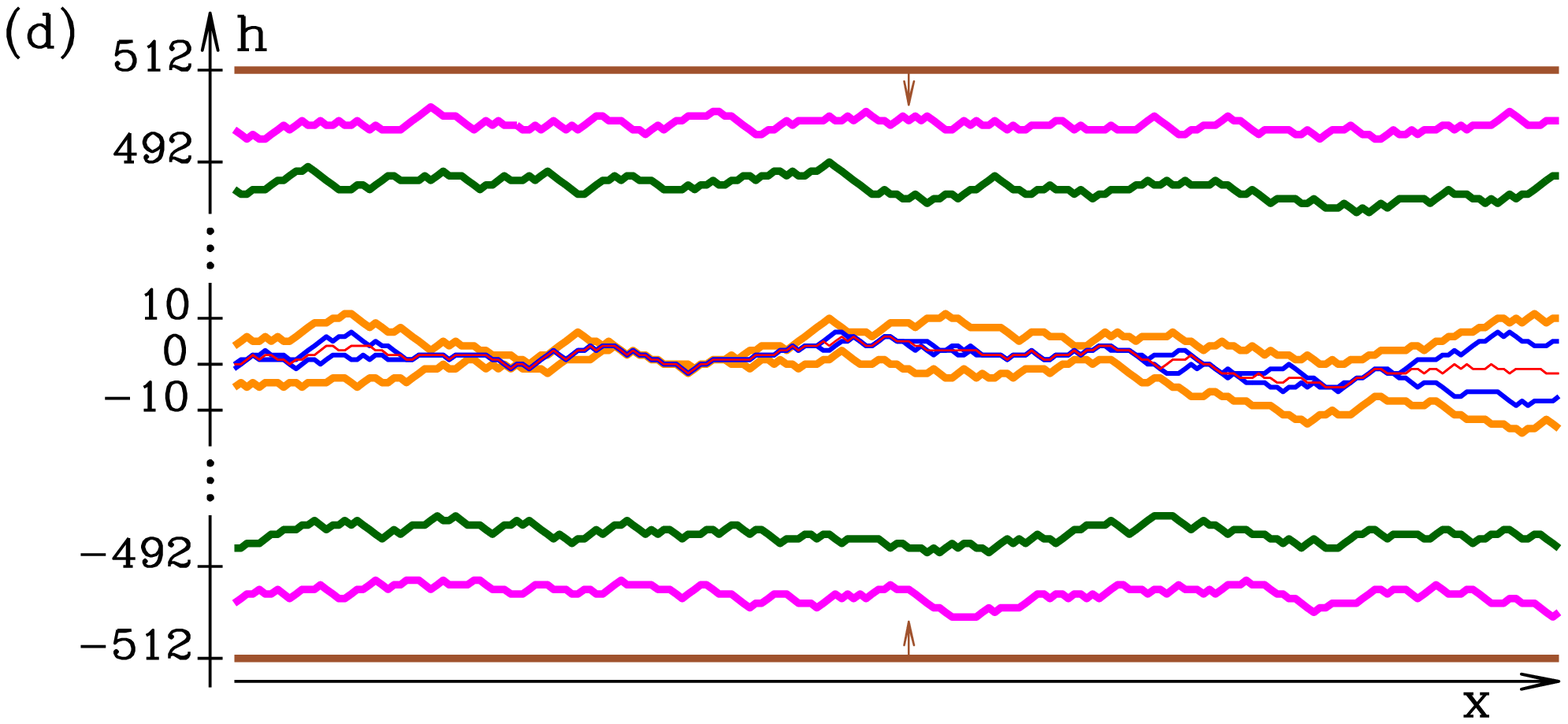}
\end{minipage}

\caption{
Interface collision models.
(a) Growth models.
Random Deposition (RD) model: at each time-step, 
an incident particle sticks with probability $p$ in each column.
Modified Family model: the incident particle aggregates at the column of incidence if no
nearest neighbor (NN) column has smaller height; if only one NN has a smaller height, it
aggregates at that column, and if two NN columns have smaller
heights, one of them is randomly chosen.
RSOS models: the particle sticks only if the 
resulting differences of heights between all NN columns do not exceed 1
(RSOS model) or 2 (RSOS2 model).
(b) Schematics of collision with short range interaction, with growth ceasing
at columns $0$, $3$, and $7$.
(c) Phantom collision, where
growth continued in columns $0$ (advance of
upper interface) and $3$ (advance of lower interface).
(d) Collision simulation ($d_0=2^9$) with two interfaces growing with the RSOS model
in opposite directions and colliding with short range interaction.
}
\label{models}
\end{figure*}

We performed simulations using well-known
one-dimensional irreversible lattice growth models: random deposition (RD)
with a sticking coefficient~\cite{Barabasi1995},
a modified Family model~\cite{family},
and restricted solid on solid models~\cite{kk}
with maximum height differences 1 (RSOS) or 2 (RSOS2).
Their rules are described in Fig.~\ref{models}(a).
The lattice constant is the unit length and the interface length is denoted as $L$.
The unit time is set by $L$ attempts of particle deposition;
rejection of such attempts are possible in RSOS and RSOS2 models or after
collision events with short range interactions (defined below).


We denote the two interface positions at time $t$ and 
abscissa $x$ as $h_-\left( x,t\right)$ and $h_+\left( x,t\right)$.
They are initially flat and located at positions $h_\pm( x,t=0)=\pm d_0$.
During growth, these interfaces move toward each other and 
collide. At each $x$ the collision time $t_c(x)$ and
the locus of the collision $h_c(x)$ obey
\begin{eqnarray}
h_+\left( x,t_c\left( x\right) \right)=h_-\left( x,t_c\left( x\right) \right)=h_c\left( x\right) .
\label{e:collision_equation}
\end{eqnarray}
Since we consider irreversible growth models, 
interfaces only move forward and, consequently,
they only pass one time at a given height.
Thus, $t_c(x)$ and $h_c(x)$ are uniquely defined
by Eq.(\ref{e:collision_equation}).
Collisions are studied when both interfaces are in their growth regimes, i. e. with
time increasing roughness \cite{Barabasi1995,Krug1997}.

The growth models are supplemented with rules describing the interaction
of the interfaces as they collide.
The first rule, which is illustrated in Fig. \ref{models}(b), accounts in a simple way
for short range interactions:
the interfaces stop growing at each column $x$ when
they meet,  i.e. when Eq.(\ref{e:collision_equation}) is satisfied.
Since particle deposition depends on the height of neighboring sites (except in RD),
the collision at a given column affects the subsequent growth of its neighbors.
An example of the dynamics with short range interaction is
presented in Fig.~\ref{models}(d).
The second rule considers non-interacting interfaces which continue to grow 
as if the opposite interface was not there.
This rule, hereafter denoted as
phantom collision, is illustrated in Fig. \ref{models}(c)
(movies of collisions with both types of rules are
reported as Supplemental Material).


We assume that interfaces move with constant and model-dependent average
velocities $v_\pm$
~\footnote{Subdominant terms in the scaling behavior are known
to affect front velocities with a slowly varying function~\cite{tiagoKPZ1d2013}.
These effects are negligible for the largest $d_0$ studied here.}.
We have $v_\pm=p_\pm$  in RD, and $v=1$ in the Family model by construction.
Moreover, we extracted from simulations
$v=0.41904(10)$ for RSOS, and $v=0.6036(3)$ for RSOS2.
The relative velocity of the two interfaces is $\bar v=v_-+v_+$, 
leading to the average collision time $t_0=2d_0/\bar v$,
while the average position of the collision is $h_0=d_0(v_--v_+)/\bar v$.
The deviations of $t_c(x)$ and $h_c(x)$ from these average values are denoted as
\begin{subequations}
\label{e:collision_deviations}
\begin{eqnarray}
\delta t_c(x)&=&t_c(x)-t_0,
\\
\delta h_c(x)&=&h_c(x)-h_0.
\end{eqnarray}
\end{subequations}
We then define the distributions $F_c(\delta t_c)$ of collision times,
and $P_c(\delta h_c)$ 
of collision loci for an initial distance $2d_0$.

\begin{figure*}
\includegraphics[width=17cm]{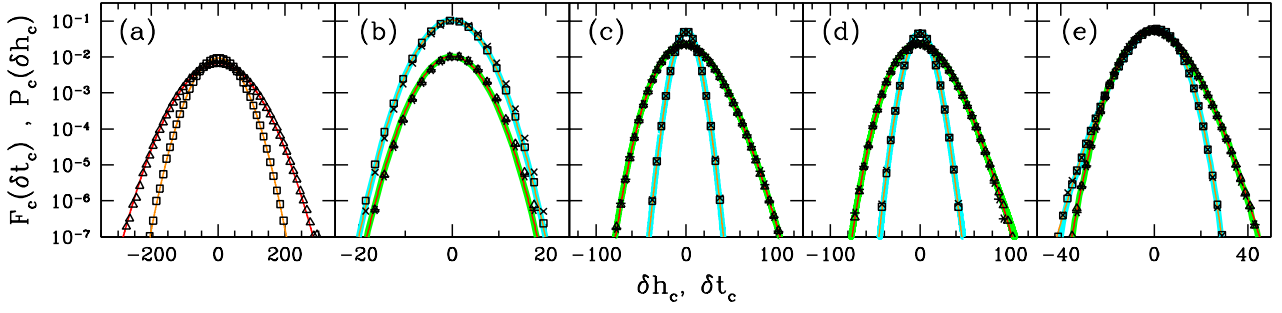}
\caption{Distributions of collision times and loci.  
Models:
(a) RD-RD with $p_+=0.5$ and $p_-=1$,
(b) Family-Family,
(c) RSOS-RSOS,
(d) RSOS-RSOS2,
(e) RSOS-Family.
Symbols indicate simulation results with
$d_0=2^{12}$ and $L=2^{32}$, and with
short range interaction [$\square$~: $P_c(\delta h_c)$,  $\triangle$~: $F_c(\delta t_c)$]
or phantom collision [$\times$~: $P_c(\delta h_c)$, $*$~: $F_c(\delta t_c)$].
Thick curves are  obtained from Eqs. (\ref{e:distribution}a,b) (green, blue)
using distributions from interfaces without collision calculated numerically.
Thin lines are distributions obtained from Eqs. (\ref{e:distribution}a,b) (red, orange)
using theoretical universal distributions (Gaussian or Tracy-Widom) with variances
extracted from simulations of interfaces without collisions.
In (b), $F_c(\delta t_c)$ is shifted down by one unit for the sake of clarity.
}
\label{dist}
\end{figure*}

The first striking point revealed by simulations is the irrelevance
of short-range interactions on the statistical properties of the collisions.
Indeed, for $d_0$ large enough, the distributions 
$F_c(\delta t_c)$ and $P_c(\delta h_c)$ in phantom collisions 
are found to be identical to those with short range interactions.
This is shown in Fig.~\ref{dist} for collision between interfaces
governed by identical or different models.


This result suggests that interactions during collision are irrelevant.
We thus define the distributions $P_\pm(\zeta_\pm;t)$
of interface fluctuations  $\zeta_\pm\left( x,t\right)=\mp\left[ h_\pm\left( x,t\right)-
\langle h_\pm\left( x,t\right) \rangle\right]$
in absence of collision
(with this definition $\zeta>0$ for fluctuations in the direction of growth).
Assuming that interactions are irrelevant,
we replace interface fluctuations by  $\zeta_\pm$,
and rewrite Eq. (\ref{e:collision_deviations}) using
 Eq. (\ref{e:collision_equation}):
\begin{subequations}
\label{e:delta_c}
\begin{eqnarray}
\delta t_c(x)=-\frac{\zeta_+(x,t_0+\delta t_c(x))+\zeta_-(x,t_0+\delta t_c(x))}{\bar v},
\label{e:delta_tc}
\\
\delta h_c(x)=\frac{-v_-\zeta_+(x,t_0+\delta t_c(x))+v_+\zeta_-(x,t_0+\delta t_c(x))}{\bar v}.
\label{e:delta_hc}
\end{eqnarray}
\end{subequations}
For large $t_0$, we expect
$\delta t_c(x)\ll t_0$, and hence to leading order we  
approximate  $t_0+\delta t_c(x)$  by $t_0$ in the r.h.s. 
of Eqs.(\ref{e:delta_c}).
We therefore define
\begin{subequations}
\label{e:delta_h0}
\begin{eqnarray}
\delta t_c^0(x)&=&-\frac{\zeta_+(x,t_0)+\zeta_-(x,t_0)}{\bar v},
\label{e:delta_tc0}
\\
\delta h_c^0(x)&=&\frac{-v_-\zeta_+(x,t_0)+v_+\zeta_-(x,t_0)}{\bar v}.
\label{e:delta_hc0}
\end{eqnarray}
\end{subequations}
These quantities
can be obtained as follows:
(i) perform the evolution as if interfaces could evolve
and freely cross without interacting up to time $t_0$;
(ii) freeze the interfaces at $t=t_0$ and slide them (forward and backward in time) 
without shape change and with their own average velocity $v_\pm$;
(iii) measure the collision times $t_c^0(x)$ and locations $h_c^0(x)$. 
This process, hereafter referred to as the freeze-and-slide approximation,
corresponds to a situation where fluctuations 
during collision are absent.

Since $\zeta_+$ and $\zeta_-$ are independent,
the probability distributions
resulting from  Eq.(\ref{e:delta_h0}) read:
\begin{subequations}
\label{e:distribution}
\begin{eqnarray}
F_c\left(\delta t_c\right)&=& \bar v\int d\zeta_+ P_+\left( \zeta_+;t_0\right)
P_-\left( -\delta t_c\bar v-\zeta_+;t_0\right)
\label{e:distribution_t_c}
\\
P_c\left( \delta h_c\right)&=&\frac{\bar v}{v_+}\int d\zeta_+ 
P_+\left( \zeta_+;t_0\right) P_-\left( \frac{-\delta h_c \bar v+\zeta_+v_-}{v_+};t_0\right).
\label{e:distribution_h_c}
\end{eqnarray}
\end{subequations}
Using $P_\pm\left( \zeta_\pm,t_0\right)$ 
obtained numerically from simulations of interfaces without collision,
we calculated these convoluted distributions for
collisions with five pairs of models, as shown in Fig.~\ref{dist}.
In all cases, there is excellent agreement with
distributions obtained in collision simulations,
confirming the validity of the freeze-and-slide approximation.

Based on this result, we now show that collision properties obey
simple scaling laws.
From dynamic scaling~\cite{Vicsek1992,Barabasi1995}, 
time correlation functions  of a growing interface
are characterized by the growth exponent $\beta$:
\begin{eqnarray}
\langle {\left[\zeta\left( x,t+\tau\right) -\zeta\left( x,t\right)\right]}^2\rangle
=B|\tau|^{2\beta} ,
\label{e:correl_time}
\end{eqnarray}
as long as the correlation length $\xi_{corr}\sim t^{\beta/\alpha}$
is smaller than the interface length $L$.
The roughness exponent $\alpha$ characterizes spatial correlations  
at short enough distances $\xi\ll\xi_{corr}$~\cite{Barabasi1995,Krug1997} via
\begin{eqnarray}
\langle {\left[\zeta\left( x+\xi,t\right) -\zeta\left( x,t\right)\right]}^2\rangle=A|\xi|^{2\alpha} .
\label{e:correl_space}
\end{eqnarray}
Within this description, RD corresponds to diffusive dynamics with $\beta=1/2$
without lateral correlation.
The other models belong to universality classes
with subdiffusive time-correlations~\cite{Barabasi1995,Krug1997}:
Edwards-Wilkinson (EW) class  with $\beta=1/4$ and $\alpha=1/2$ for 
the Family model; Kardar-Parisi-Zhang (KPZ) class
with $\beta=1/3$ and $\alpha=1/2$ for RSOS and RSOS2.


The variances of the distributions $F_c(\delta t_c)$ and $P_c(\delta h_c)$
are obtained  from Eq.(\ref{e:delta_h0}) as
\begin{subequations}
\label{e:delta_c^0_scaling_beta}
\begin{eqnarray}
\langle\delta t_c(x)^2\rangle &=&
\frac{B_+t_0^{2\beta_+}+B_-t_0^{2\beta_-}}
{\bar v^2},
\label{e:delta_t_c^0_scaling_beta}
\\
\langle\delta h_c(x)^2\rangle &=& 
\frac{v_-^2B_+t_0^{2\beta_+}+v_+^2B_-t_0^{2\beta_-}}
{\bar v^2} .
\label{e:delta_h_c^0_scaling_beta}
\end{eqnarray}
\end{subequations}
where we have used that $\langle \zeta_\pm(x,t_0)^2\rangle = B_\pm t_0^{2\beta_\pm}$
from Eq.(\ref{e:correl_time}) with $\zeta_\pm(x,t=0)=0$.
If $\beta_+=\beta_-$, both terms in the r.h.s. of Eqs.(\ref{e:delta_c^0_scaling_beta})
are equally relevant. Otherwise, for $\beta_+\neq \beta_-$, the term with the largest exponent
is asymptotically dominant, and
the variances scale with exponent $2\beta_m$, 
where $m=+$ when $\beta_+\geq\beta_-$ and $m=-$ when $\beta_->\beta_+$.

In collisions with the RD model,
each column is equivalent to an independent first passage processes,
thereby providing an alternative analytical
derivation of Eqs.(\ref{e:distribution},\ref{e:delta_c^0_scaling_beta}) in a special case.
The resulting distribution
for the height $h_\pm$ of one column is a binomial distribution~\cite{Barabasi1995}.
Using Stirling's formula, one obtains a Gaussian distribution for $P(h_\pm;t)$
at long times 
with variance $\langle \zeta_\pm ^2\rangle=\langle (h_\pm-v_\pm t)^2\rangle=4D_\pm t$,
where $v_\pm=p_\pm$ and $D_\pm=p_\pm(1-p_\pm)/2$ is the diffusion constant. 
Comparison with Eq.(\ref{e:correl_time}) leads to
$\beta_\pm=1/2$ and $B_\pm=2D_\pm$.
In one column, the collision then reduces to the first passage process
of two particles undergoing biased diffusion toward each other,
which has a well known solution~\cite{Redner2001}.
Since columns are independent, the average over realizations
leads to the same result as the average over the interface size $L$,
providing the distributions $F_c(\delta t_c)$ and $P_c(\delta h_c)$
(detailed expressions are in the Supplemental Material).
In the limit where $d_0\gg 1$ and $d_0\gg D_\pm/v_\pm$,
one finds Gaussians in agreement with Eqs.(\ref{e:distribution}),
with  variances given by Eqs.(\ref{e:delta_c^0_scaling_beta}).


\begin{table}

\begin{tabular}{|c|c|c|c|c|}
\hline
+  & Family & RSOS    & RSOS & RSOS
\\
-  & Family & RSOS    & RSOS2  & Family 
\\
\hhline{|=|=|=|=|=|}
 $\beta$($\delta t_c$) &  0.248(5)  &0.329(4)      &  0.333(1)     &  0.330(15)   
\\
Eq.(\ref{e:delta_t_c^0_scaling_beta})    &  1/4    &1/3      &  1/3       & 1/3      
\\
\hline
$\langle \delta t_c^2\rangle/(2d_0)^{2\beta}$ &  0.159(1) & 0.815(2) & 0.761(1) &  0.095(25) 
\\ 
  Eq.(\ref{e:delta_t_c^0_scaling_beta}) & 0.1589(4) &  0.814(3)& 0.759(4) &  0.0999(4) 
\\
\hline
$A_{\delta t_c}$ &  0.318(7)& 2.34(2)  &  3.6(2) & 0.75(3)
\\
 Eq.(\ref{e:correl_space_tc})   &  0.320(5) & 2.35(3)  & 3.49(7) &  0.728(10) 
\\
\hline
$\beta(\delta h_c)$  &  0.250(2) &  0.333(1) & 0.334(1)  &  0.327(3)
\\
 Eq.(\ref{e:delta_h_c^0_scaling_beta}) &  1/4  & 1/3  & 1/3  &  1/3
\\
\hline
$\langle \delta h_c^2\rangle/(2d_0)^{2\beta}$ &  0.1593(3)  & 0.1435(5)  & 0.179(1)  & 0.100(5)
\\
Eq.(\ref{e:delta_h_c^0_scaling_beta}) &  0.1589(4) &  0.1429(6) &  0.1785(6) &  0.0999(4)
\\
\hline
$A_{\delta h_c}$ &  0.320(5)& 0.414(6) & 0.745(15) &  0.475(20)
\\
 Eq.(\ref{e:correl_space_hc}) & 0.320(5) &  0.413(5)&  0.761(14)&  0.466(5)
\\
\hline
\end{tabular}

\caption{Comparison of exponents and amplitudes calculated in simulations
with short range interaction (upper values)
and predicted by the freeze-and-slide approximation
(lower values). 
}
\label{t:table1}
\end{table}


For collisions with other models, the estimates of the exponents of the variances
of $\delta t_c$ and $\delta h_c$
were obtained in simulations and are shown in Table \ref{t:table1}
(numerical procedures are in the Supplemental Material).
They agree with the exponent $\beta_m$ expected
from Eq.(\ref{e:delta_c^0_scaling_beta}).
Using the theoretically predicted value of $\beta_m$
and the variances from simulations, we calculated
the ratios $\langle \delta t_c^2\rangle/(2d_0)^{2\beta_m}$
and $\langle \delta h_c^2\rangle/(2d_0)^{2\beta_m}$ and extrapolated them to $d_0\to\infty$.
As shown in Table \ref{t:table1}, the results
agree with the  estimates obtained from
Eq.(\ref{e:delta_c^0_scaling_beta})
with the values of $v$ and $B$  
extracted from  simulations of interfaces
in the absence of collision
[$B=0.4495(10)$ for the Family model;
$B=0.254(1)$ for RSOS; 
$B=0.552(2)$ for RSOS2].


Beyond exponents, the different universality classes
impose that $P(\zeta;t)=f(\zeta/W)/W$, with $W=B^{1/2}t^\beta$ and
universal distributions $f$ at long times: Gaussian
for RD and EW class, and  Tracy-Widom for the KPZ class~\cite{Sasamoto2010,TW}.
Inserting this ansatz into Eq.(\ref{e:distribution})
and using the variances from the corresponding models without collision
at $t_0$, we obtain distributions $F_c$ and $P_c$ in
good agreement with collision simulations, as shown in Fig.~\ref{dist}
(this is confirmed by the analysis of the
skewness and kurtosis in the Supplemental Material).
If $\beta_+\neq\beta_-$,
this scaling ansatz can be inserted in Eqs.(\ref{e:distribution}).
We then find that, to leading order, 
the distributions of time and locus of collision
 follow the universal distribution of the 
growing interface with  exponent $\beta_m$: 
$F_c(\delta t_c)=f_m(-\delta t_c/T_c)/T_c$,
where $T_c=W_m/\bar v$,  and $P_c(\delta h_c)=f_m(-\delta t_c/W_c)/W_c$,
where $W_c=v_{-m}W_m/\bar v$. 
In contrast, when $\beta_+=\beta_-$ the distributions $P_c$ and $F_c$
resulting from Eq.(\ref{e:delta_c^0_scaling_beta}) cannot be rescaled
by a single time or lengthscale; they are non universal in the sense that they depend
on (ratios of) non-universal model-dependent parameters ($v_\pm$ and $B_\pm$).


We now turn to spatial correlations.
Approximating $\delta t_c$ and $\delta h_c$ by
Eq.(\ref{e:delta_h0}) and using Eq. (\ref{e:correl_space}),  we find that
at distances smaller than the correlation lengths of the two interfaces,
spatial correlations obey
\begin{subequations}
\label{e:correl_space_c}
\begin{eqnarray}
\langle {\left[\delta t_c\left( x+\xi\right)-\delta t_c\left( x\right)\right]}^2\rangle=
\frac{A_+|\xi|^{2\alpha_+}+A_-|\xi|^{2\alpha_-}}{\bar v^2},
\label{e:correl_space_tc}
\\
\langle {\left[\delta h_c\left( x+\xi\right)-\delta h_c\left( x\right)\right]}^2\rangle=
\frac{v_-^2A_+|\xi|^{2\alpha_+}+v_+^2A_-|\xi|^{2\alpha_-}}{\bar v^2}.
\label{e:correl_space_hc}
\end{eqnarray}
\end{subequations}
Thus, to leading order, correlations scale 
in $\xi$ with an exponent $\alpha_c=\max[\alpha_+,\alpha_-]$. 

In the absence of collisions,
the scaling in Eq.(\ref{e:correl_space}) is 
observed numerically in narrow ranges of $\xi$ even at long times.
However, using the known values of $\alpha_\pm$
and an extension of the procedure developed in \cite{chamereis2004},
we estimated the amplitudes $A=0.64(1)$ for the Family model,
$A=0.825(10)$ for RSOS, and $A=2.82(6)$ for RSOS2.
The same method is used to estimate
$A_{\delta t_c}\equiv\langle {\left[\delta t_c\left( x+\xi\right) -\delta t_c\left( x\right)\right]}^2\rangle /|\xi|^{2\alpha_c}$
and $A_{\delta h_c}\equiv\langle {\left[\delta h_c\left( x+\xi\right)-\delta h_c\left( x\right)\right]}^2\rangle /|\xi|^{2\alpha_c}$.
The results shown in Table \ref{t:table1} indicate good
agreement between Eq.(\ref{e:correl_space_c}) and the simulations
(the convergence to these values is presented in the Supplemental Material).
For Family-RSOS collisions, observe that $\alpha_+=\alpha_-$,
thus EW correlations contribute to the lateral correlation of the collision interface
at small lenghtscales,
although  distributions $F_c$ and $P_c$ belong to the  KPZ class.


In addition, dynamic scaling  provides a rationale for the irrelevance of 
short-range interactions.
Indeed, from Eqs.(\ref{e:delta_c^0_scaling_beta}), the collision duration
$T_c=\langle\delta {t_c\left( x\right)}^2\rangle^{1/2}\sim W_c/{\bar v}$, where 
$W_c=\langle\delta {h_c\left( x\right)}^2\rangle^{1/2}$. 
Thus, during collision, lateral correlations propagate on a distance
$\xi_{coll}\sim  T_c^{\beta/\alpha}\sim  W_c^{\beta/\alpha}$
(here the indexes of $\alpha$ and $\beta$
can be $+$ or $-$ without affecting the conclusions).  
Since the distance between the interfaces during collision
is $\sim W_m\sim W_c$, 
we expect the typical distance between contact points
to be $\xi_{contact}\sim W_c^{1/\alpha}$ from Eq.(\ref{e:correl_space_hc}).
For normal dynamic scaling, $\beta<\alpha\leq 1$\cite{Barabasi1995,Krug1997}, 
thus we have $\xi_{coll}\ll W_c \leq \xi_{contact}$
at long times.
Hence, interactions influence the collisions in the vicinity of contact points,
but these perturbations do not have time to propagate 
between contact points during the collision time.
Thus, interactions are irrelevant to leading order.

Scaling also imposes the irrelevance of fluctuations during
collision.
Indeed, we have  $T_c\sim W_c/\bar v\sim t_0^{\beta_m} \ll t_0$,
justifying the separation of scales at the origin 
of the freeze-and-slide approximation.
Furthermore, from Eqs. (\ref{e:delta_c},\ref{e:delta_h0}) 
and Eq.(\ref{e:correl_time}), we have
$\langle{\left(\delta t_c-\delta t_c^0\right)}^2\rangle
\sim\langle {\left[\zeta\left( t_0+\delta t_c\right) -\zeta\left( t_0\right)\right]}^2\rangle\sim T_c^{2\beta_m}\sim t_0^{2\beta_m^2}$.
Thus, $\langle{\left(\delta t_c-\delta t_c^0\right)}^2\rangle \ll T_c^2\sim t_0^{2\beta_m}$.
This means that deviations of $\delta t_c$ from $\delta t_c^0$
are negligible, i.e. fluctuations during collision are irrelevant.
This result and a similar analysis of
$\langle{\left(\delta h_c-\delta h_c^0\right)}^2\rangle$ 
are presented in the Supplemental Material. 
Similarly, when the growing fronts reach the late-times stationary state
where the roughness saturates to a value that depends on L,
scaling as a function of L is also expected for large L, as observed in simulations
in Refs.\cite{Albano1997,Albano2001}.


As a final remark, we conjecture that our results
for irreversible growth should directly extend to
growing interfaces with particle attachment and detachment,
that may exhibit more than one passage 
obeying Eq.(\ref{e:collision_equation}). 
In such cases, the predictions reported above describe the average
passage time for phantom collisions instead of their first passage time.
Nevertheless, the difference between the first passage time
and the average passage time is dictated by
the fluctuations during the collision, which were shown to be irrelevant.
As a consequence, the first passage time should also be 
well approximated by the 
freeze and slide process and our results should be valid
 when backward motion of the interfaces is  possible.
This conclusion is corroborated by the 
agreement discussed above between the asymptotic behaviors of the
irreversible RD model and the continuum biased random walk,
which exhibits both forward and backward propagation.


In conclusion, our central result is that local interactions 
and interface fluctuations during the collision do
not affect the asymptotic statistical properties of interface collision.
As a consequence, collision properties exhibit dynamic scaling with universal exponents;
however, distributions can be non-universal when  $\beta_+=\beta_-$.

Our results may be investigated with the measurement of grain boundary roughness
of two-dimensional materials such as
graphene~\cite{Gao2010,Huang2011,Yu2011,kiraly2013}
and MoS${}_2$~\cite{tao2017,karvonen2017}.
Assume for example that the radius $R$ of growing two-dimensional grains
is proportional to time $t$, and 
$\beta$ is the growth exponent of the
two grain edges before collision.
From Eq.(\ref{e:delta_h_c^0_scaling_beta}), we speculate that the roughness
of grain boundaries will be $W\sim t^\beta\sim R^\beta$.
The relation between $W$ and $R$ 
should therefore allow one to determine $\beta$,
providing strong constraints on the possible microscopic growth mechanisms
proposed in the literature\cite{Logovina2008,Wu2015}.

As a promising perspective, interface collisions can be considered as a
generalization of first passage processes~\cite{Redner2001,Metzler2014},
where particles diffuse and stick or annihilate when they meet. 
As opposed to particles, interfaces present intrinsic roughness, 
which leads to a spreading of the collision in time
(some parts meet earlier than others)
and in space (all parts do not meet on the same plane).
Hence, advances
on first-passage of subdiffusive systems~\cite{Metzler2014,Guerin2016}
and in exact solutions of kinetic roughening~\cite{Sasamoto2010,Calabrese2011,DeNardis2017}
should provide tools to explore the underlying links between interface collisions
and first passage processes.
Natural ramifications linked to
persistence~\cite{Bray2013},
large deviation \cite{meersonPRL2016},
and extremal statistics\cite{dentzPRE2016} of interfaces, 
also appear when e.g. considering the properties of first and last
contacts during interface collisions.


\begin{acknowledgments}

FDAAR acknowledges support by CNPq and FAPERJ (Brazilian agencies)
and thanks the hospitality of Universit\'e Lyon 1, where part of this work was performed.
OPL wishes to thank Nanoheal (EU H2020 research and innovation program under grant agreement No \textbf{642976}),
and LOTUS (ANR-13-BS04-0004-02 Grant).

\end{acknowledgments}


\begin{thebibliography}{40}%
\makeatletter
\providecommand \@ifxundefined [1]{%
 \@ifx{#1\undefined}
}%
\providecommand \@ifnum [1]{%
 \ifnum #1\expandafter \@firstoftwo
 \else \expandafter \@secondoftwo
 \fi
}%
\providecommand \@ifx [1]{%
 \ifx #1\expandafter \@firstoftwo
 \else \expandafter \@secondoftwo
 \fi
}%
\providecommand \natexlab [1]{#1}%
\providecommand \enquote  [1]{``#1''}%
\providecommand \bibnamefont  [1]{#1}%
\providecommand \bibfnamefont [1]{#1}%
\providecommand \citenamefont [1]{#1}%
\providecommand \href@noop [0]{\@secondoftwo}%
\providecommand \href [0]{\begingroup \@sanitize@url \@href}%
\providecommand \@href[1]{\@@startlink{#1}\@@href}%
\providecommand \@@href[1]{\endgroup#1\@@endlink}%
\providecommand \@sanitize@url [0]{\catcode `\\12\catcode `\$12\catcode
  `\&12\catcode `\#12\catcode `\^12\catcode `\_12\catcode `\%12\relax}%
\providecommand \@@startlink[1]{}%
\providecommand \@@endlink[0]{}%
\providecommand \url  [0]{\begingroup\@sanitize@url \@url }%
\providecommand \@url [1]{\endgroup\@href {#1}{\urlprefix }}%
\providecommand \urlprefix  [0]{URL }%
\providecommand \Eprint [0]{\href }%
\providecommand \doibase [0]{http://dx.doi.org/}%
\providecommand \selectlanguage [0]{\@gobble}%
\providecommand \bibinfo  [0]{\@secondoftwo}%
\providecommand \bibfield  [0]{\@secondoftwo}%
\providecommand \translation [1]{[#1]}%
\providecommand \BibitemOpen [0]{}%
\providecommand \bibitemStop [0]{}%
\providecommand \bibitemNoStop [0]{.\EOS\space}%
\providecommand \EOS [0]{\spacefactor3000\relax}%
\providecommand \BibitemShut  [1]{\csname bibitem#1\endcsname}%
\let\auto@bib@innerbib\@empty
\bibitem [{\citenamefont {Gao}\ \emph {et~al.}(2010)\citenamefont {Gao},
  \citenamefont {Guest},\ and\ \citenamefont {Guisinger}}]{Gao2010}%
  \BibitemOpen
  \bibfield  {author} {\bibinfo {author} {\bibfnamefont {L.}~\bibnamefont
  {Gao}}, \bibinfo {author} {\bibfnamefont {J.~R.}\ \bibnamefont {Guest}}, \
  and\ \bibinfo {author} {\bibfnamefont {N.~P.}\ \bibnamefont {Guisinger}},\
  }\href {\doibase 10.1021/nl1016706} {\bibfield  {journal} {\bibinfo
  {journal} {Nano Letters}\ }\textbf {\bibinfo {volume} {10}},\ \bibinfo
  {pages} {3512} (\bibinfo {year} {2010})},\ \bibinfo {note} {pMID: 20677798},\
  \Eprint {http://arxiv.org/abs/http://dx.doi.org/10.1021/nl1016706}
  {http://dx.doi.org/10.1021/nl1016706} \BibitemShut {NoStop}%
\bibitem [{\citenamefont {Huang}\ \emph {et~al.}(2011)\citenamefont {Huang},
  \citenamefont {Ruiz-Vargas}, \citenamefont {van~der Zande}, \citenamefont
  {Whitney}, \citenamefont {Levendorf}, \citenamefont {Kevek}, \citenamefont
  {Garg}, \citenamefont {Alden}, \citenamefont {Hustedt}, \citenamefont {Zhu},
  \citenamefont {Park}, \citenamefont {McEuen},\ and\ \citenamefont
  {Muller}}]{Huang2011}%
  \BibitemOpen
  \bibfield  {author} {\bibinfo {author} {\bibfnamefont {P.~Y.}\ \bibnamefont
  {Huang}}, \bibinfo {author} {\bibfnamefont {C.~S.}\ \bibnamefont
  {Ruiz-Vargas}}, \bibinfo {author} {\bibfnamefont {A.~M.}\ \bibnamefont
  {van~der Zande}}, \bibinfo {author} {\bibfnamefont {W.~S.}\ \bibnamefont
  {Whitney}}, \bibinfo {author} {\bibfnamefont {M.~P.}\ \bibnamefont
  {Levendorf}}, \bibinfo {author} {\bibfnamefont {J.~W.}\ \bibnamefont
  {Kevek}}, \bibinfo {author} {\bibfnamefont {S.}~\bibnamefont {Garg}},
  \bibinfo {author} {\bibfnamefont {J.~S.}\ \bibnamefont {Alden}}, \bibinfo
  {author} {\bibfnamefont {C.~J.}\ \bibnamefont {Hustedt}}, \bibinfo {author}
  {\bibfnamefont {Y.}~\bibnamefont {Zhu}}, \bibinfo {author} {\bibfnamefont
  {J.}~\bibnamefont {Park}}, \bibinfo {author} {\bibfnamefont {P.~L.}\
  \bibnamefont {McEuen}}, \ and\ \bibinfo {author} {\bibfnamefont {D.~A.}\
  \bibnamefont {Muller}},\ }\href {\doibase 10.1038/nature09718} {\bibfield
  {journal} {\bibinfo  {journal} {Nature}\ }\textbf {\bibinfo {volume} {469}},\
  \bibinfo {pages} {389} (\bibinfo {year} {2011})}\BibitemShut {NoStop}%
\bibitem [{\citenamefont {Yu}\ \emph {et~al.}(2011)\citenamefont {Yu},
  \citenamefont {Jauregui}, \citenamefont {Wu}, \citenamefont {Colby},
  \citenamefont {Tian}, \citenamefont {Su}, \citenamefont {Cao}, \citenamefont
  {Liu}, \citenamefont {Pandey}, \citenamefont {Wei}, \citenamefont {Chung},
  \citenamefont {Peng}, \citenamefont {Guisinger}, \citenamefont {Stach},
  \citenamefont {Bao}, \citenamefont {Pei},\ and\ \citenamefont
  {Chen}}]{Yu2011}%
  \BibitemOpen
  \bibfield  {author} {\bibinfo {author} {\bibfnamefont {Q.}~\bibnamefont
  {Yu}}, \bibinfo {author} {\bibfnamefont {L.~A.}\ \bibnamefont {Jauregui}},
  \bibinfo {author} {\bibfnamefont {W.}~\bibnamefont {Wu}}, \bibinfo {author}
  {\bibfnamefont {R.}~\bibnamefont {Colby}}, \bibinfo {author} {\bibfnamefont
  {J.}~\bibnamefont {Tian}}, \bibinfo {author} {\bibfnamefont {Z.}~\bibnamefont
  {Su}}, \bibinfo {author} {\bibfnamefont {H.}~\bibnamefont {Cao}}, \bibinfo
  {author} {\bibfnamefont {Z.}~\bibnamefont {Liu}}, \bibinfo {author}
  {\bibfnamefont {D.}~\bibnamefont {Pandey}}, \bibinfo {author} {\bibfnamefont
  {D.}~\bibnamefont {Wei}}, \bibinfo {author} {\bibfnamefont {T.~F.}\
  \bibnamefont {Chung}}, \bibinfo {author} {\bibfnamefont {P.}~\bibnamefont
  {Peng}}, \bibinfo {author} {\bibfnamefont {N.~P.}\ \bibnamefont {Guisinger}},
  \bibinfo {author} {\bibfnamefont {E.~A.}\ \bibnamefont {Stach}}, \bibinfo
  {author} {\bibfnamefont {J.}~\bibnamefont {Bao}}, \bibinfo {author}
  {\bibfnamefont {S.-S.}\ \bibnamefont {Pei}}, \ and\ \bibinfo {author}
  {\bibfnamefont {Y.~P.}\ \bibnamefont {Chen}},\ }\href {\doibase
  10.1038/nmat3010} {\bibfield  {journal} {\bibinfo  {journal} {Nat Mater}\
  }\textbf {\bibinfo {volume} {10}},\ \bibinfo {pages} {443} (\bibinfo {year}
  {2011})}\BibitemShut {NoStop}%
\bibitem [{\citenamefont {Kiraly}\ \emph {et~al.}(2013)\citenamefont {Kiraly},
  \citenamefont {Iski}, \citenamefont {Mannix}, \citenamefont {Fisher},
  \citenamefont {Hersam},\ and\ \citenamefont {Guisinger}}]{kiraly2013}%
  \BibitemOpen
  \bibfield  {author} {\bibinfo {author} {\bibfnamefont {B.}~\bibnamefont
  {Kiraly}}, \bibinfo {author} {\bibfnamefont {E.~B.}\ \bibnamefont {Iski}},
  \bibinfo {author} {\bibfnamefont {A.~J.}\ \bibnamefont {Mannix}}, \bibinfo
  {author} {\bibfnamefont {B.~L.}\ \bibnamefont {Fisher}}, \bibinfo {author}
  {\bibfnamefont {M.~C.}\ \bibnamefont {Hersam}}, \ and\ \bibinfo {author}
  {\bibfnamefont {N.~P.}\ \bibnamefont {Guisinger}},\ }\href {\doibase
  10.1038/ncomms3804} {\bibfield  {journal} {\bibinfo  {journal} {Nat.
  Commun.}\ }\textbf {\bibinfo {volume} {4}},\ \bibinfo {pages} {2804}
  (\bibinfo {year} {2013})}\BibitemShut {NoStop}%
\bibitem [{\citenamefont {Evans}\ \emph {et~al.}(2010)\citenamefont {Evans},
  \citenamefont {Hu},\ and\ \citenamefont {Keblinski}}]{Evans2010}%
  \BibitemOpen
  \bibfield  {author} {\bibinfo {author} {\bibfnamefont {W.~J.}\ \bibnamefont
  {Evans}}, \bibinfo {author} {\bibfnamefont {L.}~\bibnamefont {Hu}}, \ and\
  \bibinfo {author} {\bibfnamefont {P.}~\bibnamefont {Keblinski}},\ }\href
  {\doibase 10.1063/1.3435465} {\bibfield  {journal} {\bibinfo  {journal}
  {Applied Physics Letters}\ }\textbf {\bibinfo {volume} {96}},\ \bibinfo
  {pages} {203112} (\bibinfo {year} {2010})},\ \Eprint
  {http://arxiv.org/abs/http://dx.doi.org/10.1063/1.3435465}
  {http://dx.doi.org/10.1063/1.3435465} \BibitemShut {NoStop}%
\bibitem [{\citenamefont {Merabia}\ and\ \citenamefont
  {Termentzidis}(2014)}]{Merabia2014}%
  \BibitemOpen
  \bibfield  {author} {\bibinfo {author} {\bibfnamefont {S.}~\bibnamefont
  {Merabia}}\ and\ \bibinfo {author} {\bibfnamefont {K.}~\bibnamefont
  {Termentzidis}},\ }\href {\doibase 10.1103/PhysRevB.89.054309} {\bibfield
  {journal} {\bibinfo  {journal} {Phys. Rev. B}\ }\textbf {\bibinfo {volume}
  {89}},\ \bibinfo {pages} {054309} (\bibinfo {year} {2014})}\BibitemShut
  {NoStop}%
\bibitem [{\citenamefont {Be'er}\ \emph {et~al.}(2009)\citenamefont {Be'er},
  \citenamefont {Zhang}, \citenamefont {Florin}, \citenamefont {Payne},
  \citenamefont {Ben-Jacob},\ and\ \citenamefont {Swinney}}]{Beer2009}%
  \BibitemOpen
  \bibfield  {author} {\bibinfo {author} {\bibfnamefont {A.}~\bibnamefont
  {Be'er}}, \bibinfo {author} {\bibfnamefont {H.~P.}\ \bibnamefont {Zhang}},
  \bibinfo {author} {\bibfnamefont {E.-L.}\ \bibnamefont {Florin}}, \bibinfo
  {author} {\bibfnamefont {S.~M.}\ \bibnamefont {Payne}}, \bibinfo {author}
  {\bibfnamefont {E.}~\bibnamefont {Ben-Jacob}}, \ and\ \bibinfo {author}
  {\bibfnamefont {H.~L.}\ \bibnamefont {Swinney}},\ }\href {\doibase
  10.1073/pnas.0811816106} {\bibfield  {journal} {\bibinfo  {journal}
  {Proceedings of the National Academy of Sciences}\ }\textbf {\bibinfo
  {volume} {106}},\ \bibinfo {pages} {428} (\bibinfo {year} {2009})},\ \Eprint
  {http://arxiv.org/abs/http://www.pnas.org/content/106/2/428.full.pdf}
  {http://www.pnas.org/content/106/2/428.full.pdf} \BibitemShut {NoStop}%
\bibitem [{\citenamefont {Saito}\ and\ \citenamefont
  {M\"uller-Krumbhaar}(1995)}]{Saito1995}%
  \BibitemOpen
  \bibfield  {author} {\bibinfo {author} {\bibfnamefont {Y.}~\bibnamefont
  {Saito}}\ and\ \bibinfo {author} {\bibfnamefont {H.}~\bibnamefont
  {M\"uller-Krumbhaar}},\ }\href {\doibase 10.1103/PhysRevLett.74.4325}
  {\bibfield  {journal} {\bibinfo  {journal} {Phys. Rev. Lett.}\ }\textbf
  {\bibinfo {volume} {74}},\ \bibinfo {pages} {4325} (\bibinfo {year}
  {1995})}\BibitemShut {NoStop}%
\bibitem [{\citenamefont {Kuhr}\ \emph {et~al.}(2011)\citenamefont {Kuhr},
  \citenamefont {Leisner},\ and\ \citenamefont {Frey}}]{Kuhr2011}%
  \BibitemOpen
  \bibfield  {author} {\bibinfo {author} {\bibfnamefont {J.-T.}\ \bibnamefont
  {Kuhr}}, \bibinfo {author} {\bibfnamefont {M.}~\bibnamefont {Leisner}}, \
  and\ \bibinfo {author} {\bibfnamefont {E.}~\bibnamefont {Frey}},\ }\href
  {http://stacks.iop.org/1367-2630/13/i=11/a=113013} {\bibfield  {journal}
  {\bibinfo  {journal} {New Journal of Physics}\ }\textbf {\bibinfo {volume}
  {13}},\ \bibinfo {pages} {113013} (\bibinfo {year} {2011})}\BibitemShut
  {NoStop}%
\bibitem [{\citenamefont {Derrida}\ and\ \citenamefont
  {Dickman}(1991)}]{Derrida1991}%
  \BibitemOpen
  \bibfield  {author} {\bibinfo {author} {\bibfnamefont {B.}~\bibnamefont
  {Derrida}}\ and\ \bibinfo {author} {\bibfnamefont {R.}~\bibnamefont
  {Dickman}},\ }\href {http://stacks.iop.org/0305-4470/24/i=4/a=006} {\bibfield
   {journal} {\bibinfo  {journal} {Journal of Physics A: Mathematical and
  General}\ }\textbf {\bibinfo {volume} {24}},\ \bibinfo {pages} {L191}
  (\bibinfo {year} {1991})}\BibitemShut {NoStop}%
\bibitem [{\citenamefont {Albano}(1997)}]{Albano1997}%
  \BibitemOpen
  \bibfield  {author} {\bibinfo {author} {\bibfnamefont {E.~V.}\ \bibnamefont
  {Albano}},\ }\href {\doibase 10.1103/PhysRevE.56.7301} {\bibfield  {journal}
  {\bibinfo  {journal} {Phys. Rev. E}\ }\textbf {\bibinfo {volume} {56}},\
  \bibinfo {pages} {7301} (\bibinfo {year} {1997})}\BibitemShut {NoStop}%
\bibitem [{\citenamefont {Albano}\ and\ \citenamefont
  {Irurzun}(2001)}]{Albano2001}%
  \BibitemOpen
  \bibfield  {author} {\bibinfo {author} {\bibfnamefont {E.~V.}\ \bibnamefont
  {Albano}}\ and\ \bibinfo {author} {\bibfnamefont {I.~M.}\ \bibnamefont
  {Irurzun}},\ }\href {http://stacks.iop.org/0305-4470/34/i=45/a=303}
  {\bibfield  {journal} {\bibinfo  {journal} {Journal of Physics A:
  Mathematical and General}\ }\textbf {\bibinfo {volume} {34}},\ \bibinfo
  {pages} {9631} (\bibinfo {year} {2001})}\BibitemShut {NoStop}%
\bibitem [{\citenamefont {Krusin-Elbaum}\ \emph {et~al.}(2001)\citenamefont
  {Krusin-Elbaum}, \citenamefont {Shibauchi}, \citenamefont {Argyle},
  \citenamefont {Gignac},\ and\ \citenamefont {Weller}}]{KrusinElbaum2001}%
  \BibitemOpen
  \bibfield  {author} {\bibinfo {author} {\bibfnamefont {L.}~\bibnamefont
  {Krusin-Elbaum}}, \bibinfo {author} {\bibfnamefont {T.}~\bibnamefont
  {Shibauchi}}, \bibinfo {author} {\bibfnamefont {B.}~\bibnamefont {Argyle}},
  \bibinfo {author} {\bibfnamefont {L.}~\bibnamefont {Gignac}}, \ and\ \bibinfo
  {author} {\bibfnamefont {D.}~\bibnamefont {Weller}},\ }\href {\doibase
  10.1038/35068515} {\bibfield  {journal} {\bibinfo  {journal} {Nature}\
  }\textbf {\bibinfo {volume} {410}},\ \bibinfo {pages} {444} (\bibinfo {year}
  {2001})}\BibitemShut {NoStop}%
\bibitem [{\citenamefont {Atis}\ \emph {et~al.}(2015)\citenamefont {Atis},
  \citenamefont {Dubey}, \citenamefont {Salin}, \citenamefont {Talon},
  \citenamefont {Le~Doussal},\ and\ \citenamefont {Wiese}}]{Atis2015}%
  \BibitemOpen
  \bibfield  {author} {\bibinfo {author} {\bibfnamefont {S.}~\bibnamefont
  {Atis}}, \bibinfo {author} {\bibfnamefont {A.~K.}\ \bibnamefont {Dubey}},
  \bibinfo {author} {\bibfnamefont {D.}~\bibnamefont {Salin}}, \bibinfo
  {author} {\bibfnamefont {L.}~\bibnamefont {Talon}}, \bibinfo {author}
  {\bibfnamefont {P.}~\bibnamefont {Le~Doussal}}, \ and\ \bibinfo {author}
  {\bibfnamefont {K.~J.}\ \bibnamefont {Wiese}},\ }\href {\doibase
  10.1103/PhysRevLett.114.234502} {\bibfield  {journal} {\bibinfo  {journal}
  {Phys. Rev. Lett.}\ }\textbf {\bibinfo {volume} {114}},\ \bibinfo {pages}
  {234502} (\bibinfo {year} {2015})}\BibitemShut {NoStop}%
\bibitem [{\citenamefont {Takeuchi}\ and\ \citenamefont
  {Sano}(2010)}]{Takeuchi2010}%
  \BibitemOpen
  \bibfield  {author} {\bibinfo {author} {\bibfnamefont {K.~A.}\ \bibnamefont
  {Takeuchi}}\ and\ \bibinfo {author} {\bibfnamefont {M.}~\bibnamefont
  {Sano}},\ }\href@noop {} {\bibfield  {journal} {\bibinfo  {journal} {Phys.
  Rev. Lett.}\ }\textbf {\bibinfo {volume} {104}},\ \bibinfo {pages} {230601}
  (\bibinfo {year} {2010})}\BibitemShut {NoStop}%
\bibitem [{\citenamefont {Maunuksela}\ \emph {et~al.}(1997)\citenamefont
  {Maunuksela}, \citenamefont {Myllys}, \citenamefont {K\"ahk\"onen},
  \citenamefont {Timonen}, \citenamefont {Provatas}, \citenamefont {Alava},\
  and\ \citenamefont {Ala-Nissila}}]{Maunuksela1997}%
  \BibitemOpen
  \bibfield  {author} {\bibinfo {author} {\bibfnamefont {J.}~\bibnamefont
  {Maunuksela}}, \bibinfo {author} {\bibfnamefont {M.}~\bibnamefont {Myllys}},
  \bibinfo {author} {\bibfnamefont {O.-P.}\ \bibnamefont {K\"ahk\"onen}},
  \bibinfo {author} {\bibfnamefont {J.}~\bibnamefont {Timonen}}, \bibinfo
  {author} {\bibfnamefont {N.}~\bibnamefont {Provatas}}, \bibinfo {author}
  {\bibfnamefont {M.~J.}\ \bibnamefont {Alava}}, \ and\ \bibinfo {author}
  {\bibfnamefont {T.}~\bibnamefont {Ala-Nissila}},\ }\href {\doibase
  10.1103/PhysRevLett.79.1515} {\bibfield  {journal} {\bibinfo  {journal}
  {Phys. Rev. Lett.}\ }\textbf {\bibinfo {volume} {79}},\ \bibinfo {pages}
  {1515} (\bibinfo {year} {1997})}\BibitemShut {NoStop}%
\bibitem [{\citenamefont {Guisoni}\ \emph {et~al.}(2011)\citenamefont
  {Guisoni}, \citenamefont {Loscar},\ and\ \citenamefont
  {Albano}}]{Guisoni2011}%
  \BibitemOpen
  \bibfield  {author} {\bibinfo {author} {\bibfnamefont {N.}~\bibnamefont
  {Guisoni}}, \bibinfo {author} {\bibfnamefont {E.~S.}\ \bibnamefont {Loscar}},
  \ and\ \bibinfo {author} {\bibfnamefont {E.~V.}\ \bibnamefont {Albano}},\
  }\href {\doibase 10.1103/PhysRevE.83.011125} {\bibfield  {journal} {\bibinfo
  {journal} {Phys. Rev. E}\ }\textbf {\bibinfo {volume} {83}},\ \bibinfo
  {pages} {011125} (\bibinfo {year} {2011})}\BibitemShut {NoStop}%
\bibitem [{\citenamefont {Pimpinelli}\ and\ \citenamefont
  {Villain}(1998)}]{Pimpinelli1998}%
  \BibitemOpen
  \bibfield  {author} {\bibinfo {author} {\bibfnamefont {A.}~\bibnamefont
  {Pimpinelli}}\ and\ \bibinfo {author} {\bibfnamefont {J.}~\bibnamefont
  {Villain}},\ }\href@noop {} {\emph {\bibinfo {title} {Physics of Crystal
  Growth}}}\ (\bibinfo  {publisher} {Cambridge University Press},\ \bibinfo
  {year} {1998})\BibitemShut {NoStop}%
\bibitem [{\citenamefont {Redner}(2001)}]{Redner2001}%
  \BibitemOpen
  \bibfield  {author} {\bibinfo {author} {\bibfnamefont {S.}~\bibnamefont
  {Redner}},\ }\href@noop {} {\emph {\bibinfo {title} {A Guide to First-Passage
  Processes}}}\ (\bibinfo  {publisher} {Cambridge University Press, Cambridge,
  England},\ \bibinfo {year} {2001})\BibitemShut {NoStop}%
\bibitem [{\citenamefont {Metzler}\ \emph {et~al.}(2014)\citenamefont
  {Metzler}, \citenamefont {Redner}, ,\ and\ \citenamefont
  {Oshanin}}]{Metzler2014}%
  \BibitemOpen
  \bibfield  {author} {\bibinfo {author} {\bibfnamefont {R.}~\bibnamefont
  {Metzler}}, \bibinfo {author} {\bibfnamefont {S.}~\bibnamefont {Redner}}, , \
  and\ \bibinfo {author} {\bibfnamefont {G.}~\bibnamefont {Oshanin}},\
  }\href@noop {} {\emph {\bibinfo {title} {First-Passage Phenomena and Their
  Applications}}},\ Vol.~\bibinfo {volume} {35}\ (\bibinfo  {publisher} {World
  Scientific, Singapor},\ \bibinfo {year} {2014})\BibitemShut {NoStop}%
\bibitem [{\citenamefont {Barab\'asi}\ and\ \citenamefont
  {Stanley}(1996)}]{Barabasi1995}%
  \BibitemOpen
  \bibfield  {author} {\bibinfo {author} {\bibfnamefont {A.}~\bibnamefont
  {Barab\'asi}}\ and\ \bibinfo {author} {\bibfnamefont {H.}~\bibnamefont
  {Stanley}},\ }\href@noop {} {\emph {\bibinfo {title} {Fractal Concepts in
  Surface Growth}}}\ (\bibinfo  {publisher} {Cambridge University Press},\
  \bibinfo {year} {1996})\BibitemShut {NoStop}%
\bibitem [{\citenamefont {Family}(1986)}]{family}%
  \BibitemOpen
  \bibfield  {author} {\bibinfo {author} {\bibfnamefont {F.}~\bibnamefont
  {Family}},\ }\href {\doibase 10.1088/0305-4470/19/8/006} {\bibfield
  {journal} {\bibinfo  {journal} {J. Phys. A: Math. Gen.}\ }\textbf {\bibinfo
  {volume} {19}},\ \bibinfo {pages} {L441} (\bibinfo {year}
  {1986})}\BibitemShut {NoStop}%
\bibitem [{\citenamefont {Kim}\ and\ \citenamefont {Kosterlitz}(1989)}]{kk}%
  \BibitemOpen
  \bibfield  {author} {\bibinfo {author} {\bibfnamefont {J.~M.}\ \bibnamefont
  {Kim}}\ and\ \bibinfo {author} {\bibfnamefont {J.~M.}\ \bibnamefont
  {Kosterlitz}},\ }\href {\doibase 10.1103/PhysRevLett.62.2289} {\bibfield
  {journal} {\bibinfo  {journal} {Phys. Rev. Lett.}\ }\textbf {\bibinfo
  {volume} {62}},\ \bibinfo {pages} {2289} (\bibinfo {year}
  {1989})}\BibitemShut {NoStop}%
\bibitem [{Note1()}]{Note1}%
  \BibitemOpen
  \bibinfo {note} {Subdominant terms in the scaling behavior are known to
  affect front velocities with a slowly varying function~\cite
  {tiagoKPZ1d2013}. These effects are negligible for the largest $d_0$ studied
  here.}\BibitemShut {Stop}%
\bibitem [{\citenamefont {Vicsek}(1992)}]{Vicsek1992}%
  \BibitemOpen
  \bibfield  {author} {\bibinfo {author} {\bibfnamefont {T.}~\bibnamefont
  {Vicsek}},\ }\href@noop {} {\emph {\bibinfo {title} {Fractal Growth
  Phenomena}}}\ (\bibinfo  {publisher} {World Scientific, Singapore},\ \bibinfo
  {year} {1992})\BibitemShut {NoStop}%
\bibitem [{\citenamefont {Krug}(1997)}]{Krug1997}%
  \BibitemOpen
  \bibfield  {author} {\bibinfo {author} {\bibfnamefont {J.}~\bibnamefont
  {Krug}},\ }\href {\doibase 10.1080/00018739700101498} {\bibfield  {journal}
  {\bibinfo  {journal} {Advances in Physics}\ }\textbf {\bibinfo {volume}
  {46}},\ \bibinfo {pages} {139} (\bibinfo {year} {1997})},\ \Eprint
  {http://arxiv.org/abs/http://dx.doi.org/10.1080/00018739700101498}
  {http://dx.doi.org/10.1080/00018739700101498} \BibitemShut {NoStop}%
\bibitem [{\citenamefont {Sasamoto}\ and\ \citenamefont
  {Spohn}(2010)}]{Sasamoto2010}%
  \BibitemOpen
  \bibfield  {author} {\bibinfo {author} {\bibfnamefont {T.}~\bibnamefont
  {Sasamoto}}\ and\ \bibinfo {author} {\bibfnamefont {H.}~\bibnamefont
  {Spohn}},\ }\href {\doibase 10.1103/PhysRevLett.104.230602} {\bibfield
  {journal} {\bibinfo  {journal} {Phys. Rev. Lett.}\ }\textbf {\bibinfo
  {volume} {104}},\ \bibinfo {pages} {230602} (\bibinfo {year}
  {2010})}\BibitemShut {NoStop}%
\bibitem [{\citenamefont {Tracy}\ and\ \citenamefont {Widom}(1994)}]{TW}%
  \BibitemOpen
  \bibfield  {author} {\bibinfo {author} {\bibfnamefont {C.~A.}\ \bibnamefont
  {Tracy}}\ and\ \bibinfo {author} {\bibfnamefont {H.}~\bibnamefont {Widom}},\
  }\href {\doibase 10.1007/BF02100489} {\bibfield  {journal} {\bibinfo
  {journal} {Commun. Math. Phys.}\ }\textbf {\bibinfo {volume} {159}},\
  \bibinfo {pages} {151} (\bibinfo {year} {1994})}\BibitemShut {NoStop}%
\bibitem [{\citenamefont {Chame}\ and\ \citenamefont
  {Reis}(2004)}]{chamereis2004}%
  \BibitemOpen
  \bibfield  {author} {\bibinfo {author} {\bibfnamefont {A.}~\bibnamefont
  {Chame}}\ and\ \bibinfo {author} {\bibfnamefont {F.~D. A.~A.}\ \bibnamefont
  {Reis}},\ }\href@noop {} {\bibfield  {journal} {\bibinfo  {journal} {Surf.
  Sci.}\ }\textbf {\bibinfo {volume} {553}},\ \bibinfo {pages} {145} (\bibinfo
  {year} {2004})}\BibitemShut {NoStop}%
\bibitem [{\citenamefont {Tao}\ \emph {et~al.}(2017)\citenamefont {Tao},
  \citenamefont {Chen}, \citenamefont {Chen}, \citenamefont {Chen},
  \citenamefont {Gui}, \citenamefont {Chen}, \citenamefont {Li},\ and\
  \citenamefont {Xu}}]{tao2017}%
  \BibitemOpen
  \bibfield  {author} {\bibinfo {author} {\bibfnamefont {L.}~\bibnamefont
  {Tao}}, \bibinfo {author} {\bibfnamefont {K.}~\bibnamefont {Chen}}, \bibinfo
  {author} {\bibfnamefont {Z.}~\bibnamefont {Chen}}, \bibinfo {author}
  {\bibfnamefont {W.}~\bibnamefont {Chen}}, \bibinfo {author} {\bibfnamefont
  {X.}~\bibnamefont {Gui}}, \bibinfo {author} {\bibfnamefont {H.}~\bibnamefont
  {Chen}}, \bibinfo {author} {\bibfnamefont {X.}~\bibnamefont {Li}}, \ and\
  \bibinfo {author} {\bibfnamefont {J.-B.}\ \bibnamefont {Xu}},\ }\href
  {\doibase 10.1021/acsami.7b00420} {\bibfield  {journal} {\bibinfo  {journal}
  {ACS Appl. Mater. Interfaces}\ }\textbf {\bibinfo {volume} {9}},\ \bibinfo
  {pages} {12073} (\bibinfo {year} {2017})}\BibitemShut {NoStop}%
\bibitem [{\citenamefont {Karvonen}\ \emph {et~al.}(2017)\citenamefont
  {Karvonen}, \citenamefont {Saynatjoki}, \citenamefont {Huttunen},
  \citenamefont {Autere}, \citenamefont {Amirsolaimani}, \citenamefont {Li},
  \citenamefont {Norwood}, \citenamefont {Pryghambarian}, \citenamefont
  {Lipsanen}, \citenamefont {Eda}, \citenamefont {Keiu},\ and\ \citenamefont
  {Sun}}]{karvonen2017}%
  \BibitemOpen
  \bibfield  {author} {\bibinfo {author} {\bibfnamefont {L.}~\bibnamefont
  {Karvonen}}, \bibinfo {author} {\bibfnamefont {A.}~\bibnamefont
  {Saynatjoki}}, \bibinfo {author} {\bibfnamefont {M.~J.}\ \bibnamefont
  {Huttunen}}, \bibinfo {author} {\bibfnamefont {A.}~\bibnamefont {Autere}},
  \bibinfo {author} {\bibfnamefont {B.}~\bibnamefont {Amirsolaimani}}, \bibinfo
  {author} {\bibfnamefont {S.}~\bibnamefont {Li}}, \bibinfo {author}
  {\bibfnamefont {R.~A.}\ \bibnamefont {Norwood}}, \bibinfo {author}
  {\bibfnamefont {N.}~\bibnamefont {Pryghambarian}}, \bibinfo {author}
  {\bibfnamefont {H.}~\bibnamefont {Lipsanen}}, \bibinfo {author}
  {\bibfnamefont {G.}~\bibnamefont {Eda}}, \bibinfo {author} {\bibfnamefont
  {K.}~\bibnamefont {Keiu}}, \ and\ \bibinfo {author} {\bibfnamefont
  {Z.}~\bibnamefont {Sun}},\ }\href {\doibase 10.1038/ncomms15714} {\bibfield
  {journal} {\bibinfo  {journal} {Nat. Commun.}\ }\textbf {\bibinfo {volume}
  {8}},\ \bibinfo {pages} {15714} (\bibinfo {year} {2017})}\BibitemShut
  {NoStop}%
\bibitem [{\citenamefont {Loginova}\ \emph {et~al.}(2008)\citenamefont
  {Loginova}, \citenamefont {Bartelt}, \citenamefont {Feibelman},\ and\
  \citenamefont {McCarty}}]{Logovina2008}%
  \BibitemOpen
  \bibfield  {author} {\bibinfo {author} {\bibfnamefont {E.}~\bibnamefont
  {Loginova}}, \bibinfo {author} {\bibfnamefont {N.~C.}\ \bibnamefont
  {Bartelt}}, \bibinfo {author} {\bibfnamefont {P.~J.}\ \bibnamefont
  {Feibelman}}, \ and\ \bibinfo {author} {\bibfnamefont {K.~F.}\ \bibnamefont
  {McCarty}},\ }\href {http://stacks.iop.org/1367-2630/10/i=9/a=093026}
  {\bibfield  {journal} {\bibinfo  {journal} {New Journal of Physics}\ }\textbf
  {\bibinfo {volume} {10}},\ \bibinfo {pages} {093026} (\bibinfo {year}
  {2008})}\BibitemShut {NoStop}%
\bibitem [{\citenamefont {Wu}\ \emph {et~al.}(2015)\citenamefont {Wu},
  \citenamefont {Zhang}, \citenamefont {Cui}, \citenamefont {Li}, \citenamefont
  {Yang},\ and\ \citenamefont {Zhang}}]{Wu2015}%
  \BibitemOpen
  \bibfield  {author} {\bibinfo {author} {\bibfnamefont {P.}~\bibnamefont
  {Wu}}, \bibinfo {author} {\bibfnamefont {Y.}~\bibnamefont {Zhang}}, \bibinfo
  {author} {\bibfnamefont {P.}~\bibnamefont {Cui}}, \bibinfo {author}
  {\bibfnamefont {Z.}~\bibnamefont {Li}}, \bibinfo {author} {\bibfnamefont
  {J.}~\bibnamefont {Yang}}, \ and\ \bibinfo {author} {\bibfnamefont
  {Z.}~\bibnamefont {Zhang}},\ }\href {\doibase 10.1103/PhysRevLett.114.216102}
  {\bibfield  {journal} {\bibinfo  {journal} {Phys. Rev. Lett.}\ }\textbf
  {\bibinfo {volume} {114}},\ \bibinfo {pages} {216102} (\bibinfo {year}
  {2015})}\BibitemShut {NoStop}%
\bibitem [{\citenamefont {Gu{\'e}rin}\ \emph {et~al.}(2016)\citenamefont
  {Gu{\'e}rin}, \citenamefont {Levernier}, \citenamefont {B{\'e}nichou},\ and\
  \citenamefont {Voituriez}}]{Guerin2016}%
  \BibitemOpen
  \bibfield  {author} {\bibinfo {author} {\bibfnamefont {T.}~\bibnamefont
  {Gu{\'e}rin}}, \bibinfo {author} {\bibfnamefont {N.}~\bibnamefont
  {Levernier}}, \bibinfo {author} {\bibfnamefont {O.}~\bibnamefont
  {B{\'e}nichou}}, \ and\ \bibinfo {author} {\bibfnamefont {R.}~\bibnamefont
  {Voituriez}},\ }\href {http://dx.doi.org/10.1038/nature18272} {\bibfield
  {journal} {\bibinfo  {journal} {Nature}\ }\textbf {\bibinfo {volume} {534}},\
  \bibinfo {pages} {356} (\bibinfo {year} {2016})}\BibitemShut {NoStop}%
\bibitem [{\citenamefont {Calabrese}\ and\ \citenamefont
  {Le~Doussal}(2011)}]{Calabrese2011}%
  \BibitemOpen
  \bibfield  {author} {\bibinfo {author} {\bibfnamefont {P.}~\bibnamefont
  {Calabrese}}\ and\ \bibinfo {author} {\bibfnamefont {P.}~\bibnamefont
  {Le~Doussal}},\ }\href {\doibase 10.1103/PhysRevLett.106.250603} {\bibfield
  {journal} {\bibinfo  {journal} {Phys. Rev. Lett.}\ }\textbf {\bibinfo
  {volume} {106}},\ \bibinfo {pages} {250603} (\bibinfo {year}
  {2011})}\BibitemShut {NoStop}%
\bibitem [{\citenamefont {De~Nardis}\ \emph {et~al.}(2017)\citenamefont
  {De~Nardis}, \citenamefont {Le~Doussal},\ and\ \citenamefont
  {Takeuchi}}]{DeNardis2017}%
  \BibitemOpen
  \bibfield  {author} {\bibinfo {author} {\bibfnamefont {J.}~\bibnamefont
  {De~Nardis}}, \bibinfo {author} {\bibfnamefont {P.}~\bibnamefont
  {Le~Doussal}}, \ and\ \bibinfo {author} {\bibfnamefont {K.~A.}\ \bibnamefont
  {Takeuchi}},\ }\href {\doibase 10.1103/PhysRevLett.118.125701} {\bibfield
  {journal} {\bibinfo  {journal} {Phys. Rev. Lett.}\ }\textbf {\bibinfo
  {volume} {118}},\ \bibinfo {pages} {125701} (\bibinfo {year}
  {2017})}\BibitemShut {NoStop}%
\bibitem [{\citenamefont {Bray}\ \emph {et~al.}(2013)\citenamefont {Bray},
  \citenamefont {Majumdar},\ and\ \citenamefont {Schehr}}]{Bray2013}%
  \BibitemOpen
  \bibfield  {author} {\bibinfo {author} {\bibfnamefont {A.~J.}\ \bibnamefont
  {Bray}}, \bibinfo {author} {\bibfnamefont {S.~N.}\ \bibnamefont {Majumdar}},
  \ and\ \bibinfo {author} {\bibfnamefont {G.}~\bibnamefont {Schehr}},\ }\href
  {\doibase 10.1080/00018732.2013.803819} {\bibfield  {journal} {\bibinfo
  {journal} {Advances in Physics}\ }\textbf {\bibinfo {volume} {62}},\ \bibinfo
  {pages} {225} (\bibinfo {year} {2013})},\ \Eprint
  {http://arxiv.org/abs/http://dx.doi.org/10.1080/00018732.2013.803819}
  {http://dx.doi.org/10.1080/00018732.2013.803819} \BibitemShut {NoStop}%
\bibitem [{\citenamefont {Meerson}\ \emph {et~al.}(2016)\citenamefont
  {Meerson}, \citenamefont {Katzav},\ and\ \citenamefont
  {Vilenkin}}]{meersonPRL2016}%
  \BibitemOpen
  \bibfield  {author} {\bibinfo {author} {\bibfnamefont {B.}~\bibnamefont
  {Meerson}}, \bibinfo {author} {\bibfnamefont {E.}~\bibnamefont {Katzav}}, \
  and\ \bibinfo {author} {\bibfnamefont {A.}~\bibnamefont {Vilenkin}},\ }\href
  {\doibase 10.1103/PhysRevLett.116.070601} {\bibfield  {journal} {\bibinfo
  {journal} {Phys. Rev. Lett.}\ }\textbf {\bibinfo {volume} {116}},\ \bibinfo
  {pages} {070601} (\bibinfo {year} {2016})}\BibitemShut {NoStop}%
\bibitem [{\citenamefont {Dentz}\ \emph {et~al.}(2016)\citenamefont {Dentz},
  \citenamefont {Neuweiler}, \citenamefont {M\'eheust},\ and\ \citenamefont
  {Tartakovsky}}]{dentzPRE2016}%
  \BibitemOpen
  \bibfield  {author} {\bibinfo {author} {\bibfnamefont {M.}~\bibnamefont
  {Dentz}}, \bibinfo {author} {\bibfnamefont {I.}~\bibnamefont {Neuweiler}},
  \bibinfo {author} {\bibfnamefont {Y.}~\bibnamefont {M\'eheust}}, \ and\
  \bibinfo {author} {\bibfnamefont {D.~M.}\ \bibnamefont {Tartakovsky}},\
  }\href {\doibase 10.1103/PhysRevE.94.052802} {\bibfield  {journal} {\bibinfo
  {journal} {Phys. Rev. E}\ }\textbf {\bibinfo {volume} {94}},\ \bibinfo
  {pages} {052802} (\bibinfo {year} {2016})}\BibitemShut {NoStop}%
\bibitem [{\citenamefont {Alves}\ \emph {et~al.}(2013)\citenamefont {Alves},
  \citenamefont {Oliveira},\ and\ \citenamefont {Ferreira}}]{tiagoKPZ1d2013}%
  \BibitemOpen
  \bibfield  {author} {\bibinfo {author} {\bibfnamefont {S.~G.}\ \bibnamefont
  {Alves}}, \bibinfo {author} {\bibfnamefont {T.~J.}\ \bibnamefont {Oliveira}},
  \ and\ \bibinfo {author} {\bibfnamefont {S.~C.}\ \bibnamefont {Ferreira}},\
  }\href {\doibase 10.1088/1742-5468/2013/05/P05007} {\bibfield  {journal}
  {\bibinfo  {journal} {J. Stat. Mech.}\ ,\ \bibinfo {pages} {P05007}}
  (\bibinfo {year} {2013})}\BibitemShut {NoStop}%
\end{thebibliography}

%

\end{document}


\title{Supplemental Material: Interface collisions}
\author{F. D. A. Aar\~ao Reis$^1$ and O. Pierre-Louis$^2$}
\affiliation{$^1$ Instituto de F\'\i sica, Universidade Federal Fluminense,
Avenida Litor\^anea s/n, 24210-340 Niter\'oi RJ, Brazil}
\affiliation{$^2$ ILM, University  Lyon 1, 43 Bd du 11 novembre 1918, 69622 Villeurbanne, France}
\date{\today}

\begin{abstract}
We provide some technical details concerning:
(i) the parameters for collision movies; 
(ii) the distributions of first passage time and locus for two particles 
undergoing biased diffusion;
(iii) the numerical convergence of exponents of the width of the distributions
$F_c(\delta t_c)$ and $P_c(\delta h_c)$; 
(iv) the skewness and
kurtosis of these distributions; 
(v) the numerical convergence of the amplitude of spatial correlations;
and (vi) the analysis of the difference
between $\delta t_c,\delta h_c$ and $\delta t_c^0,\delta h_c^0$.
\end{abstract}

\maketitle

\section{ Interface collision movies}

Movie 1 shows collision of RSOS-RSOS
interfaces with short range interactions.
The system length is $L=1024$.
We used periodic boundaries, 
and initial half separation  $d_0=512$.
Movie 2 correponds to the same system 
in the phantom case.
The collision line  is in blue.


\section{Distributions of first passage times and loci
for two particles undergoing biased diffusion}

Here, we obtain the distributions of collision time
and collision locus for arbitrary values of $d_0$ within the continuous time
model of biased diffusion.
Technical details can be found in Ref. \cite{Redner2001}.

We define first the independent variables~\cite{Luckas1954}
\begin{eqnarray}
h_\Delta&=&h_+-h_-,
\\
h_\Sigma&=&\frac{D_-h_++D_+h_-}{D_++D_-}.
\end{eqnarray}
At collision when $h_+=h_-=h_c$, 
we have $h_\Delta=0$ and $h_\Sigma=h_c$.
We also define the related constants
\begin{eqnarray}
v_\Delta &=& -v_+-v_-=-\bar v,
\\
v_\Sigma &=& \frac{-v_-D_++v_+D_-}{D_++D_-},
\\
D_\Delta &=& D_++D_-,
\\
D_\Sigma &=& \frac{D_+D_-}{D_++D_-}.
\end{eqnarray}

The first passage of $h_\Delta$ at the origin,
providing the distribution of collision times
in the RD model, reads~\cite{Redner2001}
\begin{eqnarray}
F_c\left( t\right)=\frac{2d_0}{{\left(4\pi D_\Delta t^3\right)}^{1/2}}
{\rm e}^{-{\left( 2d_0+v_\Delta t\right)}^2/\left(4D_\Delta t\right)}.
\label{e:RD_history_distribution}
\end{eqnarray}
In the limit $d_0\gg 1$, this distribution
reduces to a Gaussian distribution (with $S=0$ and $Q=0$).
In this limit, the average collision time and its variance read
\begin{eqnarray}
t_0=\langle t_c\rangle&=&\frac{2d_0}{v_++v_-} ,
\\
\langle \delta t_c^2\rangle &=& \frac{4d_0 \left( D_++D_-\right)}{{\left(v_++v_-\right)}^3}.
\end{eqnarray}

The distribution of the variable $h_\Sigma$ is that of a simple biased random walk
\begin{eqnarray}
P_\Sigma\left( h_\Sigma,t\right)=\frac{1}{{\left( 4\pi D_\Sigma t\right)}^{1/2}}
{\rm e}^{-{\left(\tilde z-v_\Sigma t\right)}^2/\left( 4D_\Sigma t\right)} ,
\end{eqnarray}
where
\begin{eqnarray}
\tilde z &=& h_\Sigma-d_0 \left(\frac{D_+-D_-}{D_++D_-} \right) .
\end{eqnarray}
Since the variables $h_\Delta$ and $h_\Sigma$ are independent, we 
obtain the distribution of the locus of collision as
\begin{eqnarray}
P_c\left( h_c\right)=\int_0^\infty dt F_c\left( t\right) P_\Sigma\left( h_c,t\right) ,
\end{eqnarray}
leading to
\begin{eqnarray}
P_c\left( z\right)=
\frac{d_0 \kappa_1}{2\pi D_\Delta^{1/2}D_\Sigma^{1/2}\kappa_2}
{\rm e}^{-d_0 v_\Delta/D_\Delta +\tilde z v_\Sigma /\left( 2D_\Sigma\right)}
K_1\left[\kappa_1\kappa_2\right] ,
\nonumber \\
\end{eqnarray}
where $K_1$ is the modified Bessel function of the second kind and 
\begin{eqnarray}
 \kappa_1&=& \left(\frac{v_\Delta^2}{D_\Delta}+\frac{v_\Sigma^2}{D_\Sigma}\right)^{1/2},
\\
\kappa_2 &=& \left(\frac{d_0^2}{D_\Delta}+\frac{\tilde z^2}{D_\Sigma}\right)^{1/2}.
\end{eqnarray}
Again, in the limit $d_0\gg 1$, the distribution of the collision locus
is Gaussian with $S=0$ and $Q=0$.
The average and variance read:
\begin{eqnarray}
h_0=\langle  h_c\rangle &=& \frac{d_0 \left( v_- - v_+\right)}{v_+ + v_-},
\\
\langle \delta h_c^2\rangle &=& \frac{4d_0 \left( D_+v_-^2 + D_-v_+^2\right)}{{\left( v_+ + v_-\right)}^3}.
\end{eqnarray}


\section{Convergence of the effective exponent of the width of the distributions}

\begin{figure}
\includegraphics[height=9cm]{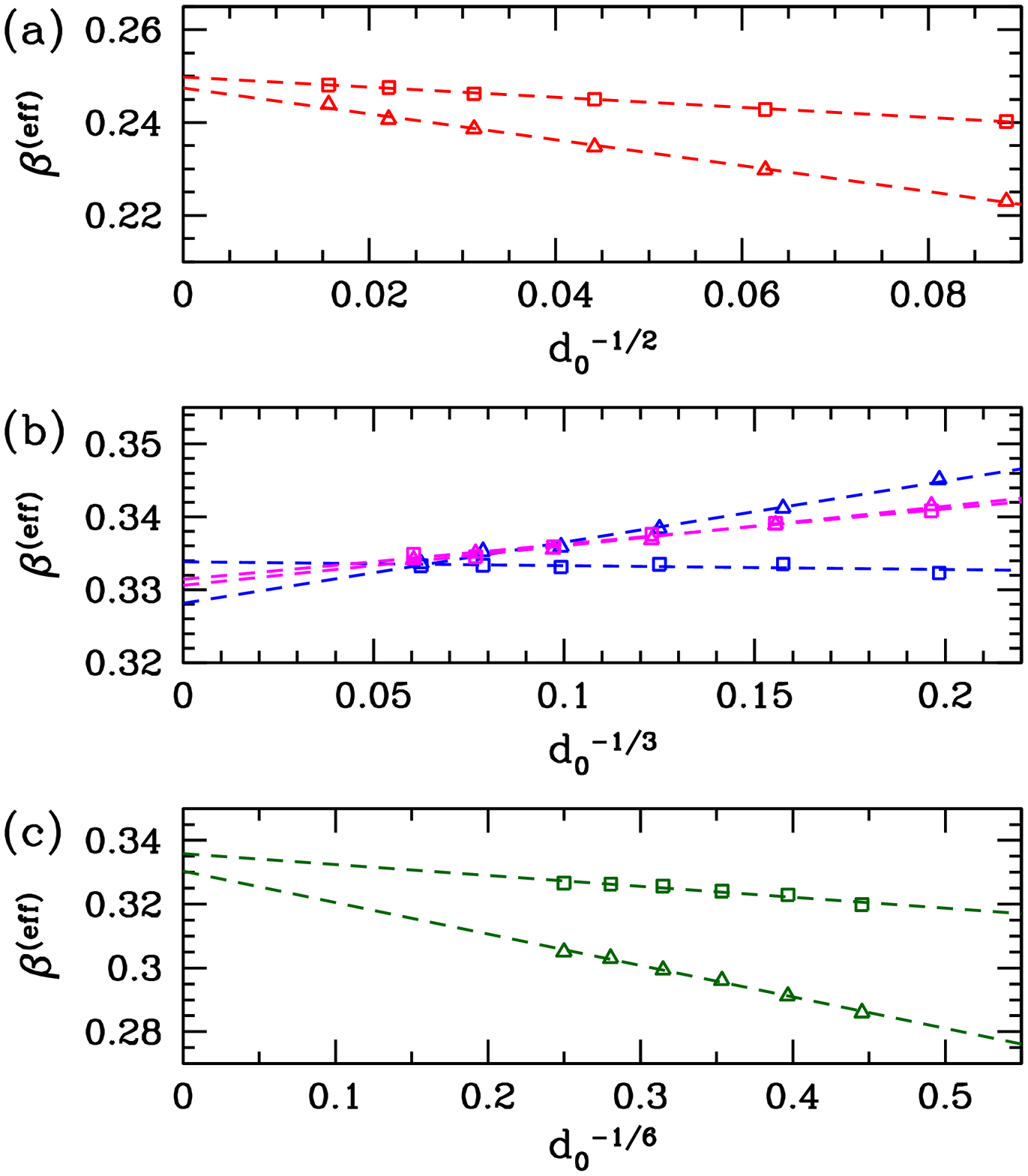}
\caption{(Color online)
Effective exponents of the variances of the distributions $F_c$ (triangles)
and $P_c$ (squares) for collisions with short range interaction:
(a) Family-Family; (b) RSOS-RSOS (blue) and RSOS-RSOS2 (magenta); (c) RSOS-Family.
Dashed lines are least squares fits of each data set.
In (c), the variable in the abscissa accounts for the correction
exponent $2\beta_- - 2\beta_+$ suggested by Eqs. (8a) and (8b) of the main text.
}
\label{beta}
\end{figure}

Effective exponents were calculated
for $\langle\delta t_c^2\rangle$ with definition
\begin{equation}
\beta^{\left( eff\right)} \left( d_0\right)\equiv 
\frac{1}{2}\frac{\ln\left[\langle\delta t_c^2( d_0)\rangle^{1/2}
/\langle\delta t_c^2( d_0/2)\rangle^{1/2}\right]}{\ln{2}} ,
\label{defbetaeff}
\end{equation}
and similarly for  $\langle\delta h_c^2\rangle$.
Their size-dependence is reported in Fig.~\ref{beta} for four collision models.
The results reported in the first and fourth rows of Table I of the main text
are obtained by the extrapolation of those plots  to $d_0\rightarrow\infty$
(linear fits shown in Fig.~\ref{beta}).

\section{Skewness and kurtosis of the distributions}

From Eqs. (4) of the main text, 
the $n$th  normalized cumulants $\kappa_n$ 
read when $\beta_+=\beta_-$~\footnote{If $\mu_n$ is the $n$-th order cumulant,
then $\kappa_n=\mu_n/\mu_2^{n/2}$.}:
\begin{subequations}
\begin{eqnarray}
\kappa_n\left(\delta t_c\right) &=& {\left(-1\right)}^n
\frac{ B_+^{n/2} \kappa_n\left(\zeta_+\right)
+B_-^{n/2}\kappa_n\left(\zeta_-\right)}
{{\left(B_++B_-\right)}^{n/2}},
\label{e:cumulants_deta_tc}
\\
\kappa_n\left(\delta h_c\right)&=&
\frac{{\left(-v_-\right)}^nB_+^{n/2} \kappa_n\left(\zeta_+\right) +v_+^nB_-^{n/2}\kappa_n\left(\zeta_-\right)}
{{\left(v_-^2B_++v_+^2B_-\right)}^{n/2}}.
\label{e:cumulants_deta_hc}
\end{eqnarray}
\label{e:cumulants_deta_c}
\end{subequations}
If $\beta_+\neq\beta_-$, the roughness of the interface with the smallest $\beta$ is negligible;
in the case $\beta_->\beta_+$ (rougher lower interface),
${\left(-1\right)}^n \kappa_n(\delta t_c)=\kappa_n(\delta h_c)=\kappa_n(\zeta_m)$;
in the case $\beta_-<\beta_+$ (rougher upper interface),
$\kappa_n(\delta t_c)=\kappa_n(\delta h_c)={\left(-1\right)}^n \kappa_n(\zeta_m)$;
The resulting predictions for the skewness $S=\kappa_3$ and the kurtosis $Q=\kappa_4$
are shown in Table S\ref{t:table1}.

The simulation values of $S$ and $Q$ are also shown in Table S\ref{t:table1}.
The results for Family-Family, RSOS-RSOS, and RSOS-RSOS2 collisions are in good agreement
with the theoretical prediction.

In the case of RSOS-Family model,
good agreement is also observed for collision locus.
However, discrepancies are observed in the results in Table S\ref{t:table1}
for the collision times.
Note that the RSOS-Family case, where one interface (Family) should be asymptotically
irrelevant as compared to the other (RSOS), 
is different from the other cases
that we have considered, where both interfaces
are relevant.
As a consequence in the RSOS-Family case, we have to reach large enough 
$d_0$ (or $t_0$) for the Family contribution to become small.
Comparing the contributions of both interfaces to the variance,
for $d_0=4096$ (the largest initial distance considered here), we find
$\langle\zeta_+^2(t_0)\rangle/\langle\zeta_-^2(t_0)\rangle\approx
\left( B_+/B_-\right) {t_0}^{2\left( \beta_+ - \beta_-\right)} \sim 0.44$;
for $d_0=128$ (smallest distance), this ratio is $\sim 0.79$.
These corrections affect the estimates of effective exponents;
in Fig. (\ref{beta}c), the value for $d_0=128$ is closer to the EW value ($0.25$)
than the KPZ value ($1/3$).
Despite this poor separation of scales, the extrapolation in  Fig. (\ref{beta}c) 
still permitted to overcome these deviations and to obtain
a  value of the exponent in agreement with the predictions,
both for $\delta t_c$ and $\delta h_c$.
However, an inspection of Fig.(\ref{beta}c) reveals
that the convergence of the properties of $\delta h_c$
are faster than the convergence of the properties of $\delta t_c$.
This convergence is actually improved by the presence of a factor $(v_+/v_-)^2$
that decreases the subdominant contribution to the variance of $\delta h_c$.
In the absence of such factor, higher order moments
of $\delta t_c$ could not be obtained
with a good accuracy.

\begin{table}
\renewcommand{\thetable}{SI}
\begin{tabular}{|c|c|c|c|c|c|c|c|}
\hline
Collision  $-$/$+$  & $S(\delta t_c)$ & $Q(\delta t_c)$&  $S(\delta h_c)$ & $Q(\delta h_c)$
\\
 &  Eq.(\ref{e:cumulants_deta_tc}) & Eq.(\ref{e:cumulants_deta_tc})&  Eq.(\ref{e:cumulants_deta_hc}) & Eq.(\ref{e:cumulants_deta_hc})
\\
\hhline{|=|=|=|=|=|}
Family/Family &  0          & 0         &  0        & 0         
\\
 & -0.005(10) & -0.005(5)  &  0.001(3) & -0.002(3) 
\\
\hline
RSOS/RSOS    &  0.2075$\dots$ & 0.0826$\dots$  & 0        & 0.0826$\dots$ 
\\
  &  0.210(5)      & 0.084(2)      &  0.000(1) & 0.082(4)      
\\
\hline
RSOS/RSOS2   & 0.2182(2) & 0.0939(2) &  0.007(2) & 0.08266(3) 
\\
  &  0.218(4)  & 0.093(5)  & 0.009(2) & 0.082(3)  
\\
\hline
RSOS/Family &   0.293$\dots$      & 0.165$\dots$       & -0.293$\dots$      & 0.165$\dots$      
\\
   & 0.22(3)           & 0.11(2)           &   -0.295(20)         & 0.17(2)           
\\
\hline
\end{tabular}
\caption{Skewness $S$ and kurtosis $Q$ of the distributions $F_c$ and $P_c$ of four collision models. The top value is the theoretical estimate 
based on the distribution of the corresponding universality class 
(Family: Gaussian, and RSOS and RSOS2: Tracy-Widom 
with\cite{Sasamoto2010,Calabrese2011} $S\approx0.2935$ and $Q\approx0.1652$).}
\label{t:table1}
\end{table}


\section{Convergence of the amplitude of spatial correlations}
\label{convergenceA}

\begin{figure}
\includegraphics[width=8cm]{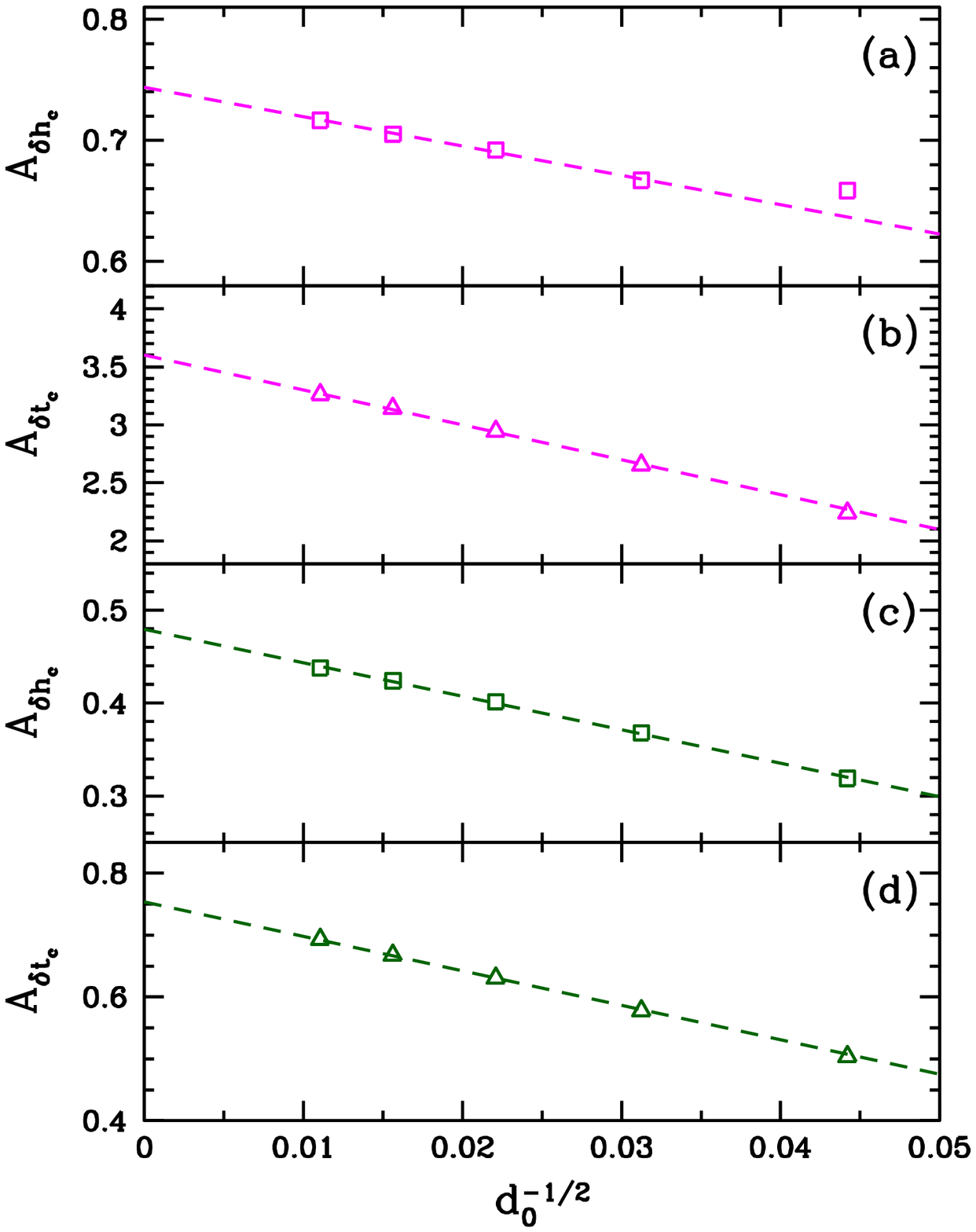}
\caption{(color online)
Amplitudes of the spatial correlation functions of the collision times (triangles)
and of the colision positions (squares) in (a),(b) RSOS-RSOS2 collisions and
(c),(d) RSOS-Family collisions.
The variable in the abscissa provides the best linear fits of the last 4 or 5
data points in each model.
}
\label{ampG2}
\end{figure}

For most values of $d_0$ considered here, the log-log plots
of $\langle (\delta h_c(x+\xi)-\delta h_c(x))^2\rangle$
and $\langle (\delta t_c(x+\xi)-\delta t_c(x))^2\rangle$ versus $|\xi|$
do not fit a straight line for small values of $|\xi|$,
even in narrow ranges of this variable.
This parallels the observations in \cite{chamereis2004} for the local roughness scaling.
For this reason, an extrapolation approach was used along the same lines of that work.
First, the ratios
$A_{\delta h_c}\left( d_0\right)\equiv\langle (\delta h_c(x+\xi)-\delta h_c(x))^2\rangle /|\xi|^{2\alpha_c}$ and
$A_{\delta t_c}\left( d_0\right)\equiv\langle (\delta t_c(x+\xi)-\delta t_c(x))^2\rangle /|\xi|^{2\alpha_c}$
were calculated for three or four fixed values of $\xi$ with the expected exact exponent $\alpha_c$;
the values of $\xi$ used here were between $8$ and $64$, so that they are not very small but
also do not exceed the lateral correlation length $\xi_{corr}$, which indicates the saturation
of those ratios.
Subsequently, those ratios were plotted as a function of ${d_0}^{-1/2}$ for each value of $\xi$;
the linear fits of those plots converge to approximately the same values in the limit
$d_0\to\infty$; this confirms the scaling
$\langle (\delta h_c(x+\xi)-\delta h_c(x))^2\rangle\sim A_{\delta h_c} |\xi|^{2\alpha_c}$
for very large $d_0$ (and equivalent for $\delta t_c$).
Fig. \ref{ampG2} illustrates the extrapolation of data with $\xi = 16$
for RSOS-RSOS2 and RSOS-Family collisions.
The asymptotic estimates for these and for the other collision models are presented in
Table I of the main text. They agree with the predictions from Eqs. (9) of the main text.
Note that both terms in  Eqs. (9a) and (9b) contribute
to the amplitudes of RSOS-Family collisions because both interface models have the same
exponent $\alpha=1/2$.

\section{Difference between $\delta h_c^0$, $\delta t_c^0$, and $\delta t_c$, $\delta h_c$}

\begin{figure}
\includegraphics[width=8cm]{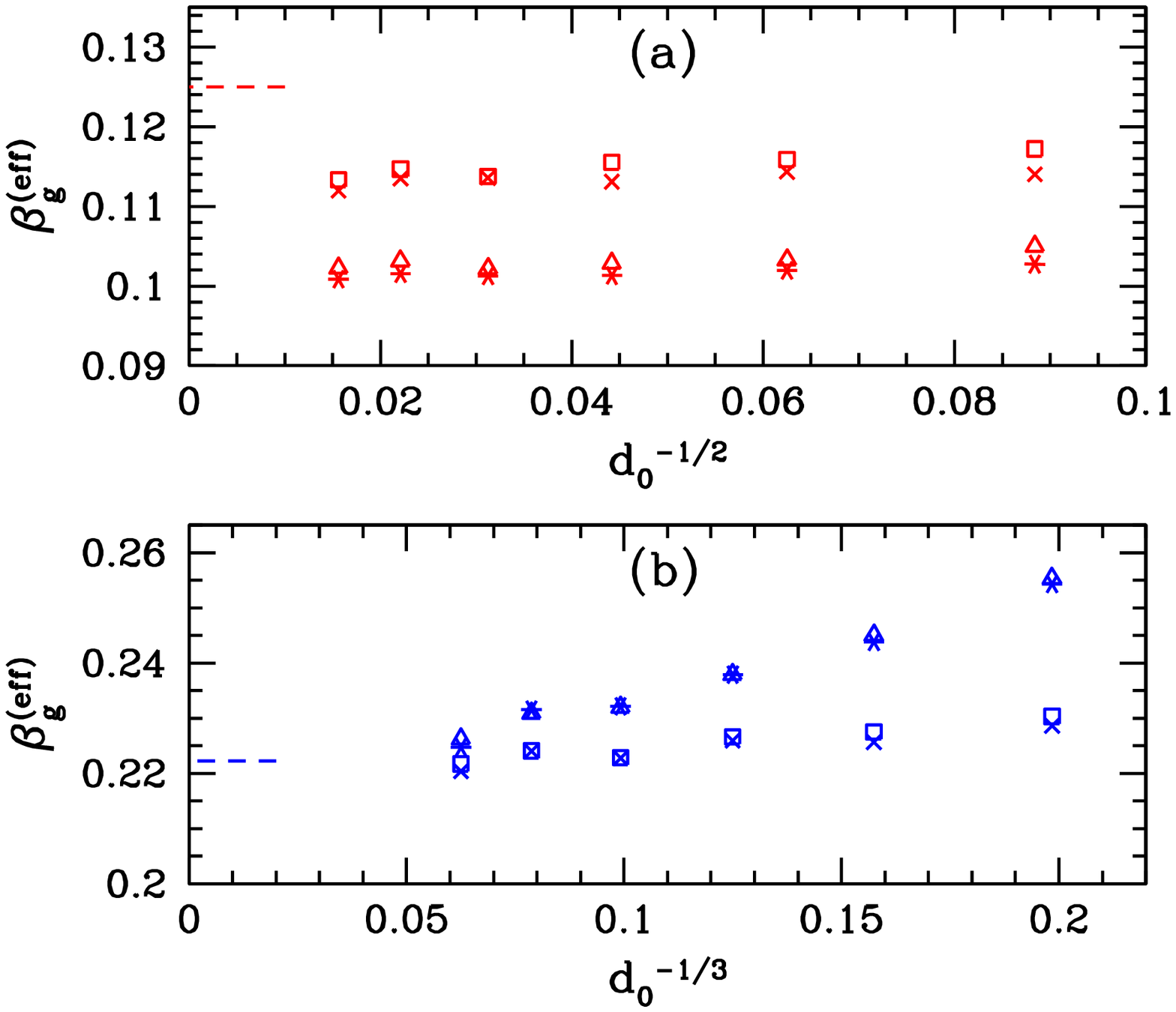}
\caption{(Color online) Effective exponents of the correlation functions
for (a) Family-Family and
(b) RSOS-RSOS collisions with short range interaction: squares $G_{\delta h_c}$, $\phi=1/1024$;
crosses $G_{\delta h_c}$, $\phi=1/512$; triangles $G_{\delta t_c}$, $\phi=1/1024$;
asterisks $G_{\delta t_c}$, $\phi=1/512$. Dashed lines indicate the theoretically predicted
exponents.
The variables in the abscissas were chosen to provide the best linear fits (not shown)
in the range of $d_0$ considered here.
}
\label{betag}
\end{figure}

The difference between  freeze-and-slide collisions and
collisions observed in simulation informs us about the influence of fluctuations
during the collision, and is quantified by the correlation functions
 \begin{subequations}
 \begin{eqnarray}
G_{\delta t_c}=\langle{\left[\delta t_c\left( x\right) -\delta t_c^0\left( x\right)\right]}^2\rangle,
 \\
G_{\delta h_c}=\langle{\left[\delta h_c\left( x\right) -\delta h_c^0\left( x\right)\right]}^2\rangle.
 \end{eqnarray}
 \label{e:defG}
 \end{subequations}
As discussed in the main text, we expect
\begin{eqnarray}
G_{\delta t_c}\sim t_0^{\beta_g} ,\; \beta_g = 2\beta_m^2 .
\label{e:Gdeltatcscaling}
\end{eqnarray}
A similar reasoning also indicates that
$G_{\delta h_c}\sim t_0^{\beta_g}\ll \langle{\delta h_c^0\left( x\right)}^2\rangle$,
and  $\langle{\delta h_c\left( x\right)}^2\rangle\approx
\langle{\delta h_c^0\left( x\right)}^2\rangle$.

In simulations with short range interactions, we considered a definition of a time
$t_0\left( \phi\right)$ at which a fraction $\phi$ 
of the sites have collided.
The values of $\phi$ are small, so that the largest part of the interface
was not affected by the collision at that time.
For a given $\phi$, effective exponents of $G_{\delta t_c}$ and $G_{\delta h_c}$ are defined
similarly to Eq.(\ref{defbetaeff}).
In Fig. \ref{betag}, we show those effective exponents for Family-Family and RSOS-RSOS
collisions considering two values of $\phi$.
The estimates of $\beta_g$ for large $d_0$ (large $t_0$) are close to the values predicted
by Eq. (\ref{e:Gdeltatcscaling}), which supports the
claim that the interface morphology exhibits negligible changes during collision
[deviations to smaller $\beta_g$ for Family-Family collisions in Fig. \ref{betag}
do not invalidate this reasoning].
Also note that the precise definition of $t_0$ (related to the value of $\phi$ in
Fig. \ref{betag}) does not affect the scalings
as long as the reference time $t_0$ is located during the collision.



%